\documentclass[
    preprintnumbers,
    amsmath,
    amssymb,
    aps]{revtex4-2}

\usepackage{graphicx}
\usepackage{dcolumn}
\usepackage{bm}
\usepackage{hyperref}
\usepackage{xcolor}

\begin{document}

\title{Efficient network exploration by means of resetting self-avoiding random walkers}

\author{Gaia Colombani$^{1,2}$}
\author{Giulia Bertagnolli$^1$}%
\author{Oriol Artime$^{2,3,4,5}$}%
\email[Corresponding author:~]{oartime@ub.edu}
\affiliation{$^1$University of Trento, Department of Mathematics, Via Sommarive, 14, 38123 Povo (TN), Italy}
\affiliation{$^2$CHuB Lab, Fondazione Bruno Kessler, Via Sommarive 18, 38123 Povo (TN), Italy}
\affiliation{$^3$Departament de Física de la Matèria Condensada, Universitat de Barcelona, 08028 Barcelona, Spain}
\affiliation{$^4$Universitat de Barcelona Institute of Complex Systems (UBICS), Universitat de Barcelona, 08028 Barcelona, Spain}
\affiliation{$^5$Universitat de les Illes Balears, 07122 Palma, Spain}
 
\date{\today}

\begin{abstract}
The self-avoiding random walk (SARW) is a stochastic process whose state variable avoids returning to previously visited states. This non-Markovian feature has turned SARWs a powerful tool for modelling a plethora of relevant aspects in network science, such as network navigability, robustness and resilience. We analytically characterize self-avoiding random walkers that evolve on complex networks and whose memory suffers stochastic resetting, that is, at each step, with a certain probability, they forget their previous trajectory and start free diffusion anew. Several out-of-equilibrium properties are addressed, such as the time-dependent position of the walker, the time-dependent degree distribution of the non-visited network and the first-passage time distribution, and its moments, to target nodes. We examine these metrics for different resetting parameters and network topologies, both synthetic and empirical, and find a good agreement with simulations in all cases. We also explore the role of resetting on network exploration and report a non-monotonic behavior of the cover time: frequent memory resets induce a global minimum in the cover time, significantly outperforming the well-known case of the pure random walk, while reset events that are too spaced apart become detrimental for the network discovery. Our results provide new insights into the profound interplay between topology and dynamics in complex networks, and shed light on the fundamental properties of SARWs in nontrivial environments.
\end{abstract}

\maketitle

\section{Introduction}

Random walks (RW), and modifications thereof, are a central model in the theory of stochastic processes and play a prominent role in the description of phenomena in a wide range of spatiotemporal scales. Some classical examples are the modelling of diffusion and transport dynamics in disordered media in physics~\cite{Kac1947, Bouchaud1990}, animal movement and foraging in ecology~\cite{viswanathan2011physics}, spread of rumors or consensus dynamics in the social sciences~\cite{sen2014sociophysics}, stock price modelling in finance~\cite{black1973pricing} and search algorithms in computer science~\cite{brin1998anatomy}, to name but a few. 

A key characteristic of random walks is their Markovian nature~\cite{van1992stochastic}. 
However, in many relevant scenarios, the future state of a system is strongly influenced by its prior states, such as in protein folding and polymer growth. 
The self-avoiding random walk (SARW) stands out as a model to address some of these non-Markovian, non-intersecting problems~\cite{fisher1966shape, madras2013self, vanderzande1998lattice}. 
From a physical point of view, the SARW is a stochastic process whose random variable evolves under the condition of not visiting previously visited states. 
Tackling the infinite memory of the SARW represents a profound mathematical challenge and, thus, in comparison with the standard random walk, few rigorous results have been established. 
It has been extensively studied as non-intersecting paths on regular lattices~\cite{lawler1980self, hammersley1991self, slade1994self, guttmann2022existence}. 
In this context, simply enumerating the number $c_n$ of self-avoiding paths of given length $n$ remains as an open problem. 
In fact, the limit $\mu \equiv \text{lim}_{n \to \infty} c_n^{1/n}$, the so-called \textit{connectivity constant}, which provides a lower bound $c_n \geq \mu^n$, is unknown for $\mathbb{Z}^d$ and it has been rigorously found to be $\sqrt{2 + \sqrt{2}}$ only recently for the hexagonal lattice~\cite{duminil2012connective}. 
Furthermore, plausible non-rigorous arguments based on spin systems abound regarding SARW's critical behavior and scaling limit, yet their mathematical proof is still lacking in many cases~\cite{slade2019self}.
 
The emergence of network science more than two decades ago has been particularly fruitful for the development of the theory of stochastic processes and statistical physics~\cite{porter2016dynamical, dorogovtsev2008critical, cimini2019statistical}. On the one hand, since most networks are embedded in infinite-dimensional spaces, they provide the backbone on which mean-field predictions can be actually tested~\cite{strogatz2022fifty}. On the other hand, networks have offered an unprecedented amount of novel domain-specific scenarios on which classical dynamical models can be deployed. This has led to the development of interesting analytical techniques, and has significantly broadened the range of impactful applications of network science, such as those in, for instance, disease containment~\cite{pastor2015epidemic}, drug repurposing~\cite{pushpakom2019drug}, influence maximization~\cite{morone2015influence}, and shock and cascade spreading~\cite{haldane2011systemic, yang2017small, artime2020abrupt}.

On networked topologies, the behavior of random walks is fairly well established~\cite{burioni2005random, Lovasz1993, masuda2017random}. On the contrary, the SARW's non-Markovian nature has hindered progress toward a comprehensive characterization of its behavior on networks. Yet, during the last years, some theoretical foothold has been gained, by addressing, for example, questions such as the connectivity constant $\mu$ and the mean length of the walks computed in several types of networks~\cite{kim2016network, herrero2003self, herrero2005kinetic, herrero2007kinetic, herrero2019self, huang2006walks}, as well as the degree distribution of the network not visited by the walker~\cite{lopez2012model, tishby2016distribution, valente2022non}. 
From an application viewpoint, SARWs have been used, for instance, to detect communities~\cite{de2018community}, to model purchase activity on signed network products~\cite{wang2019self} or to model overflow cascade spreading~\cite{valente2022non}.

Searching and navigating discrete, non-regular spaces is a salient problem in many areas of science, finding applications, for instance, in neural networks~\cite{nara1993memory}, transportation planning~\cite{cascetta1997calibrating, bonomi1984n}, website ranking~\cite{brin1998anatomy} and online social networks~\cite{pei2014searching}. It is natural to think of the SARW as an efficient strategy to explore and gather information in uncertain scenarios~\cite{cristin2021information}. 
On networks, it has been numerically shown that self-avoidance can significantly facilitate network exploration, navigability, and discovery~\cite{yang2005exploring, kim2016network}. 
Analytical results show that if a walker is not allowed to backtrack, i.e., it has just a one-step self-avoiding behavior, then the cover time can be decreased with respect to the backtracking-free case~\cite{alon2007non}. 
These results prompt us to ask whether longer non-backtracking conditions could lead to a further decrease in the cover time, and whether we can mathematically characterize the behavior of such self-avoiding walkers of finite memory. This article answers these questions in the affirmative.

\section{Self-avoiding random walks}
Let us consider a network $\mathcal{N}_0$ of size $N$ that is connected, undirected and unweighted, and that lacks degree correlations~\cite{newman2018networks}. The fraction of nodes with degree $k$ is given by the degree distribution $p_0(k)$. A self-avoiding random walk on $\mathcal{N}_0$ can be mapped to a sequence of networks $ \lbrace \mathcal{N}_t \rbrace_{t \geq 0}$ of size $N_t = N - t$, where we identify the time $t < N$ as the number of steps performed by the walker, and where the visited nodes have been sequentially removed together with their links~\cite{valente2022non}. In this way, the initial network changes at each time step until the walker reaches a degree-$0$ node and it is forced to stop because it has no available neighbors to jump to, i.e., the degree-$0$ nodes are absorbing states of the dynamics. The number of degree-$0$ nodes grows in time, so it does the probability to get blocked in one of these states.

To generalize to the case of self-avoiding random walks with finite memory, we assume that at each step the walker resets its memory with a fixed probability $r$. This has several physical motivations, such as the limitations of the cognitive capacity of a living explorer or a searching strategy programmed under low complexity constraints. 
This scenario could even represent run-and-tumble motion~\cite{malakar2018steady, mori2020universal} on networks, where self-avoiding moves are related to run events while memory resets correspond to tumbles. 
Anyway, we assume the reset event restores the accessible topology to its initial configuration $\mathcal{N}_0$, while the walker keeps its position. 
Thus, memory resetting becomes a simple yet very suitable strategy to explore a network, since it gets rid, naturally, of the undesirable effect of topological traps. Yet, once the walker is blocked in a certain node, it will need to wait there until the next resetting event occurs. 
Ideally, we would like to minimize the time spent in trapped configurations, and this motivates a quest for an optimal value of $r$ to efficiently navigate the network or, at least, for a range of values for the resetting parameter, which provide a clear advantage with respect to other searching strategies.
This approach is further supported by the observation that our memory-resetting model interpolates between the pure self-avoiding walk ($r = 0$) and the standard random walk ($r = 1$).

Stochastic resetting lends itself to be studied with renewal equations~\cite{evans2020stochastic}. Hence, we need first to characterize the reset-free process, that is, the pure self-avoiding random walk, in order to obtain the quantities of interest of the SARW with resetting. Let us start with the time-dependent degree distribution $p_t(k)$ of the networks $\lbrace \mathcal{N}_t \rbrace_{t \geq 0}$. We denote by $\langle k\rangle_t$ the mean degree of $\mathcal{N}_t$, and by
\begin{equation} \label{eq:excess-deg-t}
    q_t(k)=\dfrac{(k+1)p_t(k+1)}{\langle k\rangle_t}.
\end{equation}
its excess degree distribution~\cite{newman2001random}.
In general, the excess degree distribution is the probability of finding a node with given (excess) degree following an edge. Here, we can interpret $q_t(k)$ as the probability that the walker reaches a node of degree $k$ at time $t+1$. The number of nodes of degree $k$ at time $t$ is indicated by $N_t(k)$, and we will evaluate the difference of nodes of degree $k$ in  a single time step, $D_t(k) \equiv N_t(k) - N_{t-1}(k)$. To do this, we have to distinguish whether the walker has reached a degree-$0$ or not. 
If it has reached such a node at time $t-1$, it stops and $N_t(k)$ does not change anymore. For this case we will use the notation $D_t(\cdot|0)$ and we have $D_t(k|0)=0$ for all $k$. 
If the walker has not reached a degree-$0$ node at time $t-1$, the walk proceeds and the network does change. In this latter case, we use the notation $D_t(\cdot|0^c)$. 
Since we work with initially connected networks, the number of nodes of degree $k=0$ can only increase as the walk evolves. In particular, their variation is computed as  $D_t(0|0^c)\approx\langle r\rangle_{t-2}q_{t-1}(0)$, where $\langle r\rangle_t$ is the mean value of $q_t(k)$. The population of nodes of degree $k>0$, instead, may increase---when nodes of degree $k+1$ lose an edge incident to a visited node---or decrease during the process, since some of them lose an edge or the removed node was of degree $k$. 
These scenarios correspond, respectively, to the three terms in the following equation 
\begin{equation}\label{eq:Dt}
D_t(k|0^c)\approx\langle r\rangle_{t-2}[q_{t-1}(k)-q_{t-1}(k-1)]-q_{t-2}(k).
\end{equation}
In the end, applying the law of total probability, for $t \in \{1, \dots, N-1\}$ we obtain
\begin{align}
  & p_t(k)=\dfrac{N_{t-1}(k)+D_{t}(k)}{N_t} = q_{t-2}(0)p_{t-1}(k)+(1-q_{t-2}(0))\left[\frac{N_{t-1}(k)+D_t(k|0^c)}{N_t}\right].\label{eq:ptk-node}
\end{align}
See the Supplemental Material (SM) for more details on the derivation of Eq.~\eqref{eq:ptk-node}. The excess distribution $q_{-1}(k)$ can be chosen accordingly to the initial condition of the walker. For example, if the starting position is randomly selected, then $q_{-1}(k)=p_0(k)$. In Fig.~\ref{fig:deg_dist_er} and \ref{fig:deg_dist_sf} we verify that the analytical approximation Eq.~\eqref{eq:ptk-node} compares well with the results obtained from simulations on Erd{\H{o}}s-R{\'{e}}nyi (ER) and scale-free (SF) random networks. Interestingly, simulated SARWs on SF networks stop, on average, before their counterparts on ER networks. This is investigated in the following.

Firstly, notice that $p_t(k)$ is the conditional probability of randomly selecting a degree-$k$ node at time $t$, given that the walk proceeds up to time $t$. 
Eventually, at the last time step $p_{t=N}(k) = \delta_{k0}$, since the chances that a walker has not yet been blocked decrease as time grows.
With these observations, and using the relation~\eqref{eq:excess-deg-t} between $p_t(k)$ and $q_t(k)$, we are able to compute the average length of a pure SARW on a network. 
The length of the SARW corresponds to the first time at which the walker reaches a degree-$0$ node. 
We call this a stopping event, and its corresponding random variable, the stopping time. The latter is distributed according to the distribution
\begin{equation}
    s(t)=q_{t-1}(0)\prod_{t'=0}^{t-1}(1-q_{t'-1}(0)).
\end{equation}
The average length is, simply, the expected value of the stopping time $ \langle L\rangle=\sum_{t=0}^{N-1} ts(t)$. 
Higher-moments of the length of the walk trivially follow the same reasoning. As expected, $\langle L \rangle^{\text{SF}} < \langle L \rangle^{\text{ER}}$, with walks on ER networks displaying a higher variation in their stopping times, as can be seen in Fig.~\ref{fig:stop_dist}.

\begin{figure}
\includegraphics[width=.95\textwidth]{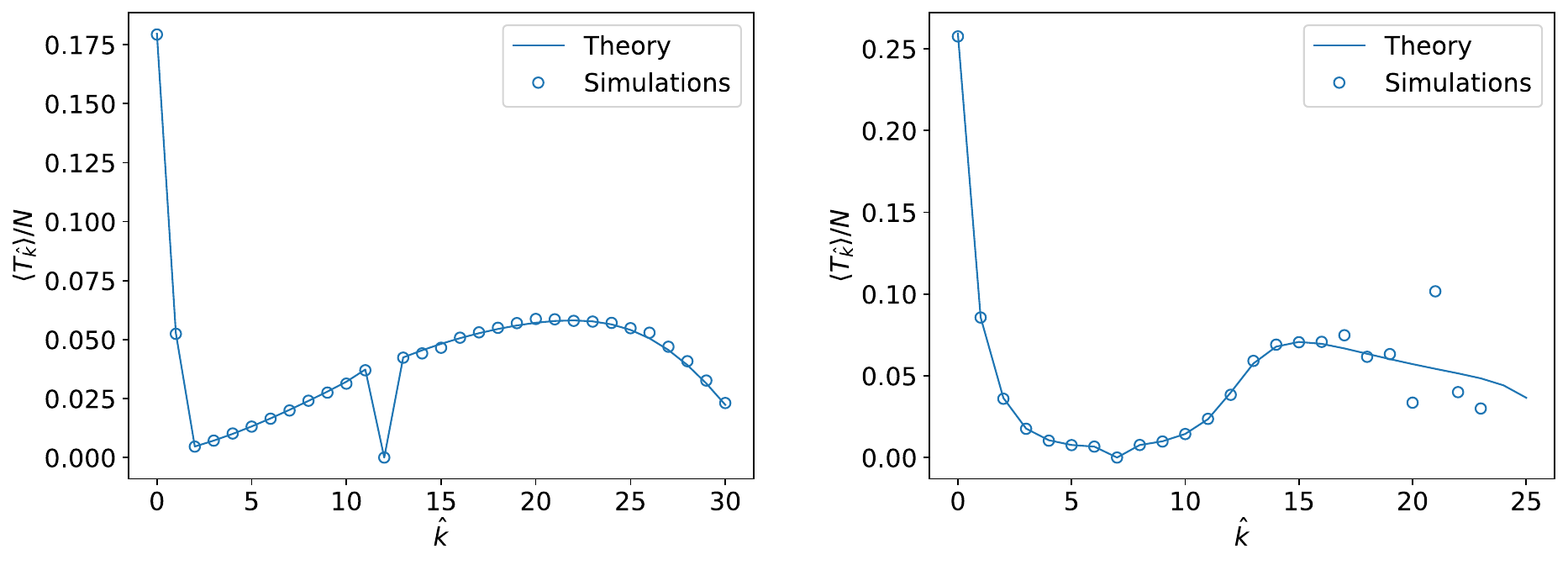}
\caption{\label{fig:mfpt_sf}Self-avoiding random walks on random graphs -- mean first-passage time to a node of degree $\hat{k}$. On two different topologies---(left) SF with $N=1000$, $\alpha=2.5$ and initial minimum degree $3$ and (right) ER with $N=1000$ and $\langle k \rangle_0 = 7$---we compare our theoretical predictions (solid lines), Eq.~\eqref{eq:mfpt}, with the simulations results (markers). Simulations results are averaged over $5000$ walker trajectories, computed on $1000$ independent SF/ER graphs. For each network, we run the SARW dynamics $5$ times, randomly choosing the initial node among those of degree $12$ (SF) or $7$ (ER). This explains the values $\langle T_{\hat{k}}\rangle = 0$ for such degrees, respectively. However, the qualitative behavior of $\langle T_{\hat{k}}\rangle$ is independent of the initial condition; see the results for arbitrary initial conditions in Figs.~\ref{fig:SM_mfpt_sf}-\ref{fig:SM_mfpt_er}.}
\end{figure}

The mean length of the SARW can be seen as a first-passage process to a node of degree $0$. We can generalize this to nodes of arbitrary degree $\hat{k} \neq 0$. To estimate the first-passage probabilities, it is necessary to consider the target degree nodes as absorbing states~\cite{van1992stochastic, artime2018first}, stopping the process when the walker reaches one of them. Hence, we modify Eq.~\eqref{eq:ptk-node}, yielding
\begin{equation}\label{eqdegdist2}
p_t(k)= \left[q_{t-2}(0)+q_{t-2}(\hat{k}) \right] p_{t-1}(k)+ (1-q_{t-2}(0)-q_{t-2}(\hat{k}))\left[\dfrac{N_{t-1}(k)+D_t(k|\{0,\hat{k}\}^c)}{N_t}\right]
\end{equation}
with $D_t(k|\{0,\hat{k}\}^c) \approx \langle r\rangle_{t-2}[q_{t-1}(k)-q_{t-1}(k-1)]$. Now the process presents two absorbing states, the degree-$0$ nodes, which are intrinsic to the SARW dynamics, and the degree-$\hat{k}$ nodes. The stopping probabilities to these states, respectively $s^{0}(t)$ and $s^{\hat{k}}(t)$, are the probabilities to reach a degree-$0$ node without reaching a degree-$\hat{k}$ node before, and vice versa. 
Thus,
\begin{eqnarray}
    && s^{0}(t)=q_{t-1}(0)\prod_{t'=0}^{t-1}(1-q_{t'-1}(0)-q_{t'-1}(\hat{k})),\label{eq:stop1}\\
    && s^{\hat{k}}(t)=q_{t-1}(\hat{k})\prod_{t'=0}^{t-1}(1-q_{t'-1}(0)-q_{t'-1}(\hat{k})).\label{eq:stop2}
\end{eqnarray}
They are not probability distributions, yet they verify $\sum_{t}\left(s^{0}(t)+s^{\hat{k}}(t)\right)=1$. Indicating the splitting probability by $\pi_{\hat{k}} = \sum_{t=0}^{N-1}s^{\hat{k}}(t)$, the mean first-passage time (MFPT) to a node of degree $\hat{k}$ is 
\begin{equation}
    \label{eq:mfpt}
    \langle T_{\hat{k}}\rangle =\frac{1}{\pi_{\hat{k}}}\sum_{t=0}^{N-1}ts^{\hat{k}}(t).
\end{equation}
If there were not the self-avoiding behavior, we would expect $\langle T_{\hat{k}}\rangle$ to be inversely proportional to $p(\hat{k})$. 
However, while the walk evolves, surviving nodes lose neighbors and, hence, the population of low-degree nodes increases at the expense of the high-degree ones. 
For this reason, the MFPT is non-monotonic as a function of $\hat{k}$, displaying a non-trivial local maximum in the intermediate-to-large degree region. 
Above that maximum, the MFPT decreases because it is difficult to step on a high-degree node, unless the walker finds it in the early exploration phase. 
Below the local maximum, the MFPT is lower for the opposite reason: those degrees accumulate in the network through the link-removal events.
If the target degree keeps decreasing, though, $\langle T_{\hat{k}}\rangle$ grows again due to the absence of nodes with very low degree in the initial walking phase. 
This pattern is clearly visible in Fig.~\ref{fig:mfpt_sf}, which shows both the theory and simulations of the mean first-passage time on scale-free and Erd{\H{o}}s-R{\'{e}}nyi networks.
Observe that this pattern does not depend on the type of network, although simulations on ER have larger fluctuations for the high values of $\hat{k}$. 
These depend on the particular degree distribution of the model, where the frequency of nodes with high degree is small, hence only for few realizations of the process, the walker effectively reaches those degrees before the walk stops.
Finally, note that Eq.~\eqref{eq:stop2} allows us to characterize the trajectories that arrive at nodes of degree $\hat{k}$ without visiting degree-$0$ nodes, i.e., that have not been trapped. This formalism can be easily generalized to devise tailored trajectories that neither visit nodes of degree $0$ nor of arbitrary degrees $\hat{k}', \hat{k}'', \ldots$, and study their out-of-equilibrium properties.

Recall that the time-dependent degree distribution $p_t(k)$, Eq.~\eqref{eq:ptk-node}, is computed under the assumption that the walker has not been blocked. So, averages are computed only over the non-trapped walks, too.
To close the analysis of the pure SARW, we focus on $p(t,k)$, the degree distribution of the network not visited by the walker, considering all walks, including those that have been trapped. 
To this end, we need to combine the conditioned degree distribution $p_t(k)$ with the stopping distribution, such that
\begin{align}
    &p(t,k)=p_t(k)\left[1-\sum_{t'=1}^{t-1}s(t') \right]+\sum_{t'=1}^{t-1}p_{t'}(k)s(t'), \label{eq:ptk-joint}
\end{align}
with the initial conditions $p(0,k)=p_0(k)$ and  $p(1,k)=p_1(k)$. 
The first term on the right-hand side is the conditioned degree distribution times the probability that the walker did not stop up to time $t$, while the summation in the second term considers all the possibilities that the walker stops exactly at any intermediate time $1\le t'< t$. In Fig.~\ref{fig:p(t,k)} we verify that the prediction~\eqref{eq:ptk-joint} reproduces well the results from simulations.
The fact that $p(t,k)$ takes into account all trajectories is evinced in its non-null stationary values, as it can be checked in Fig.~\ref{fig:p(t,k)}.

\section{Self-avoiding random walks with resetting}

Now that the reset-free self-avoiding random walk has been characterized, we are in the position to study its finite-memory version. Since the network is restored in the resetting events, the time $t$ is no longer constrained to be smaller than the system size. 
A renewal equation for $p_r(t,k)$, standing for the probability of finding a node with degree $k$ at time $t$ for the self-avoiding random walk with resetting (SARWR), can be written as 
\begin{equation}\label{eq:prtk}
    p_r(t,k)=(1-r)^{t-1}p(t,k)+\sum_{t'=1}^{t-1}(1-r)^{t'-1}r p_r(t-t',k),
\end{equation}
for all $t \geq 2$, with initial conditions $p_r(0,k)=p(0,k)$ and $p_r(1,k)=p(1,k)$. We set $p(t,k)=p(N-1,k)$ for all $t \geq N$. The first term on the right-hand side is the probability that 
no resetting events happen up to time $t$, while the second represents the probability that the first resetting occurred at any intermediate time $1\le t' <t$.

\begin{figure}
\includegraphics[width=.95\textwidth]{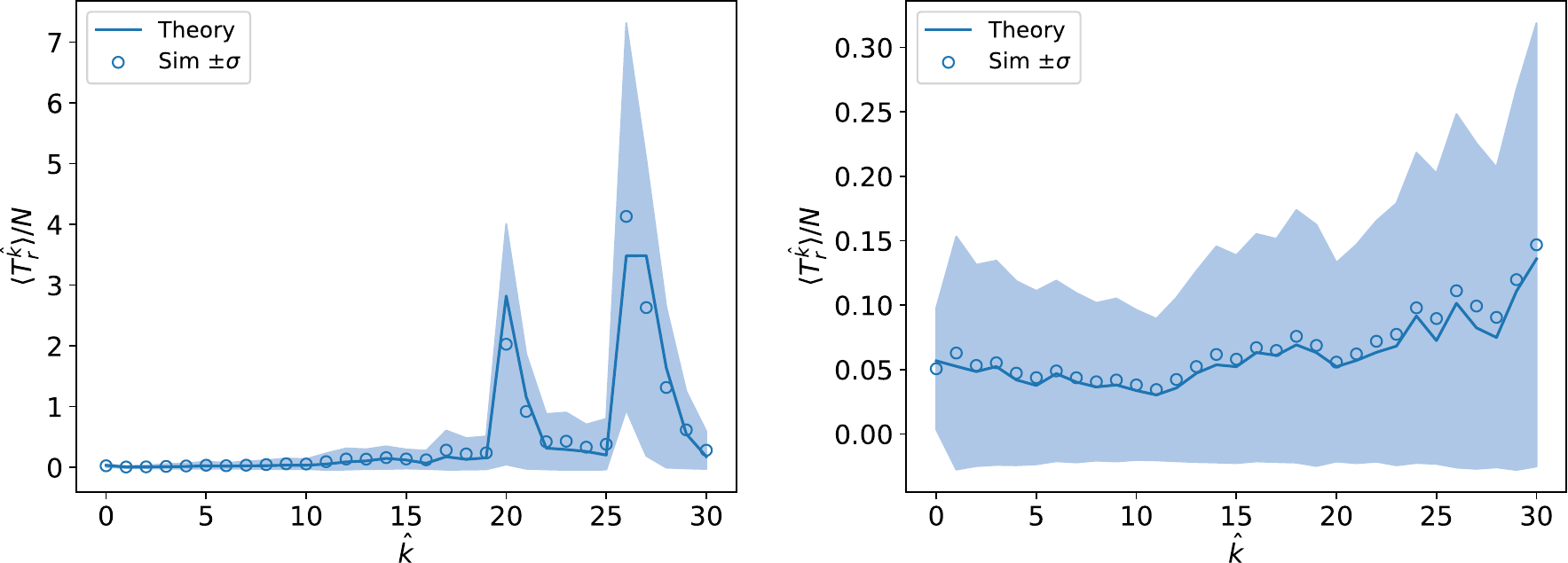}
\caption{\label{fig:mfpt_sarwr_realnets}Self-avoiding random walk with resetting on the two real networks -- mean first passage time to a node of degree $\hat{k}$. We compare our theoretical predictions (solid lines), Eq.~\eqref{eq:Trkhat}, with the simulations results (markers), for a SARWR on (right) the Federal Aviation Administration (FAA) network~\cite{kunegis2013konect,Maayan2006} and (left) on the network collecting email exchanges among members of the Rovira i Virgili University in Spain, in 2003~\cite{guimera2003self}. We have used as resetting parameter the values for which the mean waiting time between resets is approximately equal to the SARW mean length on each network, namely, $r\approx1/35$ and $r\approx1/58$, respectively. The light-blue color illustrates the standard deviation. Simulations are run for $5000$ different initial conditions, with nodes chosen uniformly at random.}
    \label{fig:email_sarwr_mfpt}
\end{figure}

Following similar steps, we evaluate first-passage quantities to a node of degree $\hat{k}$, having in mind that in this setup reaching a degree-$0$ node does not stop the walk permanently, but only until the next reset. To derive the stopping distribution, we use again a renewal equation for the survival distribution. In the reset-free process, this is simply the complementary of the cumulative stopping distribution
\begin{equation*}
    S^{\hat{k}}(t)=1-\sum_{t'=0}^{t-1}s^{\hat{k}}(t'),
\end{equation*}
with initial condition $S^{\hat{k}}(0)=1$. Then, adding the resetting, we obtain
\begin{align}
   S_r^{\hat{k}}(t)=(1-r)^{t-1}S^{\hat{k}}(t)+\sum_{t'=1}^{t-1}(1-r)^{t'-1}r S^{\hat{k}}(t'+1)S_r^{\hat{k}}(t-t')\label{eq:surv},
\end{align} 
with $S_r^{\hat{k}}(0)=S^{\hat{k}}(0)$, $S_r^{\hat{k}}(1)=S^{\hat{k}}(1)$, and $S^{\hat{k}}(t)=S^{\hat{k}}(N-1)$ whenever $t \geq N$.
The first term on the right-hand side gives the probability that, until time $t$ neither a reset nor any hitting to a degree-$\hat{k}$ have occurred.  The second term represents the probability that the first resetting occurred at some $t'$, without hitting a target neither in $(0, t')$, nor in $(t', t)$. Thus, the stopping distribution and mean first-passage time for the SARWR are given by
\begin{align}
    s_r^{\hat{k}}(t) & = S_r^{\hat{k}}(t)-S_r^{\hat{k}}(t+1) \\
    \langle T_r^{\hat{k}}\rangle & = \sum_tts^{\hat{k}}_r(t). \label{eq:Trkhat}
\end{align}
The validity of Eq.~\eqref{eq:Trkhat} has been proved performing several SARWRs on ER and SF random networks, and on two different real networks---a transportation and a communication network. The comparison between theory and simulations of the mean first-passage time of self-avoiding walkers with memory resetting on the two real networks is displayed in Figure~\ref{fig:email_sarwr_mfpt}. The results for synthetic ER and SF graphs are shown in Figs.~\ref{fig:mfpt_res_er} and \ref{fig:mfpt_res_sf}.

The real networks that we employ present homogeneous and heterogeneous connection patterns: in the transportation network the degrees are approximately exponentially distributed with scale parameter $\lambda^{-1}=9.2$, while the fitted distribution for the degrees in the communication network is a power-law with exponent $\alpha=1.9$. We study the SARWR on their largest connected component, which contain $\mathcal{O}(10^3)$ nodes. Note the good agreement of theoretical and simulated results, even if degree correlations are present, in general, in real systems.
See the SM for additional figures and details.
The addition of resetting drastically changes the behavior of the MFPT. On the one hand, the local maximum of the pure SARW disappears, since high-degree nodes are now newly accessible after each resetting. On the other hand, the resetting parameter determines two regimes: for $r \to 1$ high-degree nodes need longer time to be visited, while for $r \to 0$ low-degree nodes have larger MFPTs. 

As the last descriptor of the resetting process, we compute $d_r(t,k)$, the probability to find the walker on a degree-$k$ node at time $t$. To use the renewal approach, we need the reset-free probability $d(t,k)$ of finding the walker on a degree-$k$ node. If the walker has not stopped yet, the position is given by $q_t(k)$, otherwise it certainly has to be on a degree-$0$ node. Thus, we obtain 
\begin{align*}
    &d(t,k)=q_{t-1}(k) \left[1-\sum_{t'=1}^{t-1}s(t') \right]\quad \forall k\ne 0,
\end{align*}
and $d(t,0)=1-\sum_{k\ne0}d(t,k)$. Including memory resetting, we get
\begin{align*}
   &d_r(t,k)=(1-r)^{t-1}d(t,k)+\sum_{t'=1}^{t-1}(1-r)^{t'-1}r d_r(t-t',k),
\end{align*}
with $d_r(0,k)=d(0,k)$, $d_r(1,k)=d(1,k)$ and $d(t,k)=d(N-1,k)$ for all $t \geq N$. Remarkably, $d_r(t,k)$ captures well the semi-absorbing behavior of degree-$0$ nodes up to the resetting event. The accuracy of this equation is verified in Fig.~\ref{fig:dr(t,k)_er}, \ref{fig:dr(t,k)_sf} and \ref{fig:dr(t,k)_realnets}.

A final question that we address is the mean cover time, which is, on average, the time required to explore the entire network and, hence, quantifies the efficiency of the random search. 
The cover time for a classical random walk on a random graph, is, almost surely, less than $N^3$~\cite{aleliunas1979random}.
This strict bound is, nevertheless, much larger than the actual cover time in most networks, yet the precise mean value depends on the degree distribution~\cite{cooper2007cover, cooper2007cover_b}. For instance, in random regular graphs it drops to $N \log N$~\cite{cooper2005cover, tishby2021analytical}.
We will see next that, by combining the self-avoiding constraint with resetting, we can speed up the exploration of the network with respect to the standard RW, the pure SARW and the non-backtracking random walk. 

\begin{figure}
\includegraphics[width=.85\textwidth]{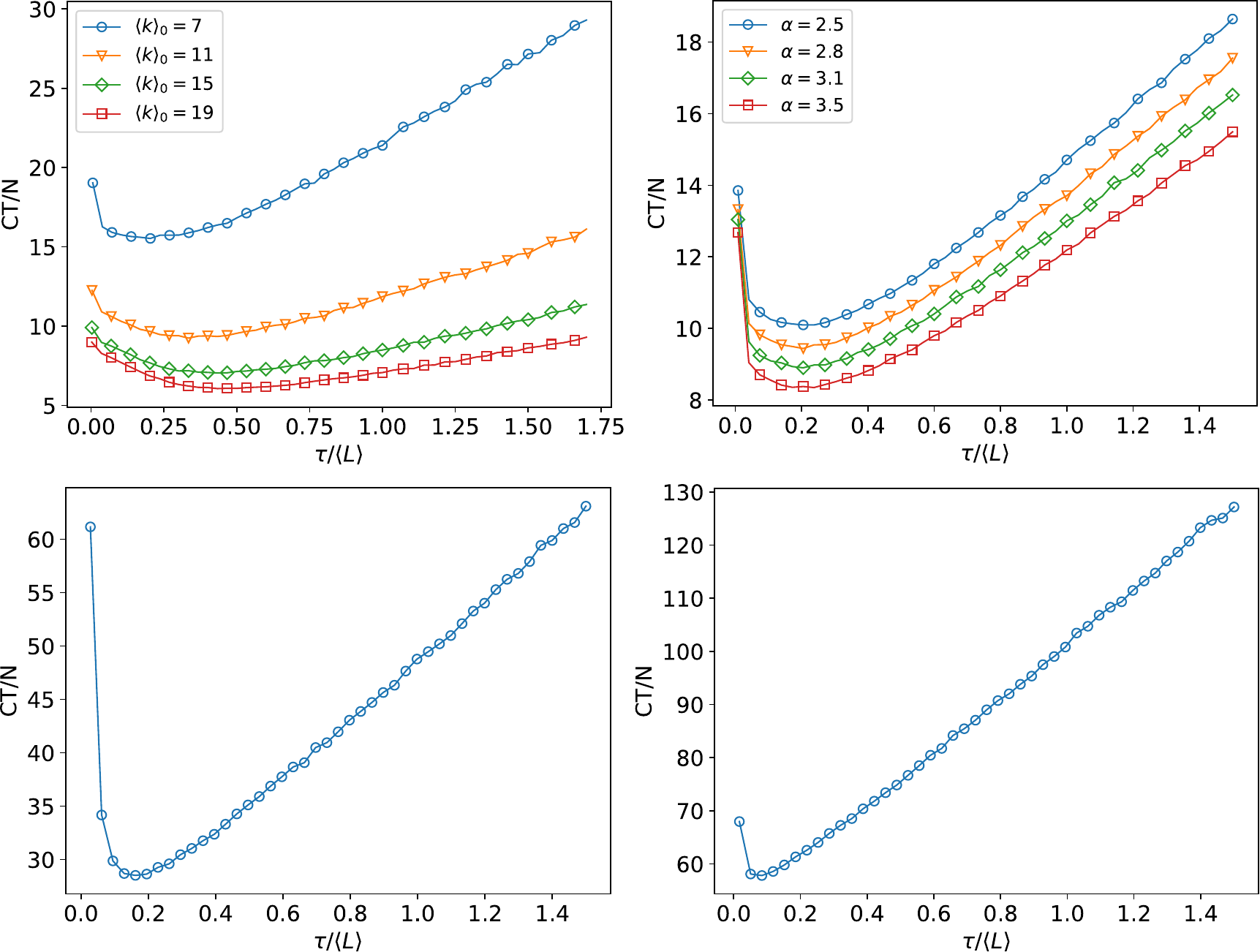}
\caption{\label{fig:cov}Mean cover time $CT$ of self-avoiding random walks with resetting for ER and SF random graphs (top row) and real networks (bottom row). The curves correspond to different values of the topological parameter of the initial degree distribution. On the $x$-axis there are the mean waiting time ($\tau=1/r$) values scaled by the mean length of the SARW in the corresponding network. For the ER cases (top left), we have $\langle L\rangle\approx \lbrace 170,\,360,\,365,\,400\rbrace$ for the increasing values of $\langle k\rangle_0$ shown in the legend. For the SF cases (top right), $\langle L\rangle\approx 120$ in all four scenarios. For the air transportation network (bottom left) $\langle L\rangle\approx35$, and for the email exchange network (bottom right) $\langle L\rangle\approx58$. The limit $\tau \to 1$ yields the classical RW case. The network size for the synthetic networks is $500$ for all cases, and averages are computed over 8000 independent realizations.}
\end{figure}

Broadly construed, the self-avoiding behavior helps the walker to escape the departing node faster than a standard random walk. 
In turn, self-avoidance blocks the walker at degree-$0$ nodes. These two effects compete in the process of network discovery. 
Intuitively, then, we can expect a nontrivial value of the resetting probability $r$ that favors long self-avoiding explorations while minimizing the time the walker is trapped in the pseudo-absorbing states. We numerically compute the cover time for the SARWR in synthetic graphs (Fig.~\ref{fig:cov}, top row) and in the two empirical systems introduced above (Fig.~\ref{fig:cov}, bottom row). We show the dependence of the mean cover time, $CT$, as a function of the mean time between resetting events $\tau \equiv 1/r$, normalized by the length of the SARW. In all the cases studied, we find the existence of an optimal value of the resetting parameter, $r^{\text{opt}}$, for which $CT$ is a global minimum.

From our analysis it emerges that the minimum cover time $CT(r^{\text{opt}})$ always occurs for $r^{\text{opt}} \neq 1$. 
Since $r=1$---implying also $\tau = 1$, i.e., minimum mean waiting time between resets---corresponds to the RW case, $r^{\text{opt}} \neq 1$ suggests that the resetting mechanism significantly boosts the exploration speed of classical random walkers.
Not any amount of resetting, however, is beneficial to expedite the search: if the resetting rate is small enough, $\tau$ increases too much and the $CT$ becomes larger than its value for the random walker case, impairing the efficiency of the search. In this setting, indeed, the walker waits too long in blocked configurations, before a resetting event allows free diffusion anew.

The optimal value the value $r^{\text{opt}}$, and the range of resetting parameters in which the memory resetting outperforms the RW search, obviously depend on the topological properties of the network. In fact, we observe that the role of the topological parameter of each network family (mean degree $\langle k \rangle$ for ER networks and exponent $\alpha$ in SF networks) impacts very differently how this cover time reduction is achieved. For SF networks, $r^{\text{opt}}(\alpha)$ is roughly the same, and the cover time curves have the same functional form, with only a slight reduction as $\alpha$ increases. Hence, the range of resetting rates for which there is a reduction is independent of $\alpha$. On the other hand, for ER, $r^{\text{opt}}(\langle k \rangle)$ increases with the mean degree, and so does the parameter region where the exploration improvement is achieved.

It is instructive to compare also the efficiency of the resetting mechanism against the case of reset-free self-avoiding walkers. One could naively think that the optimal strategy that minimizes the cover time could be to perform walks that reset their memory once they get trapped. Statistically, this would mean to set $r \approx \langle L \rangle$, where the mean length $\langle L \rangle$, of course, depends on the network, as we have shown above. The information contained in Fig.~\ref{fig:cov} reveals that this is generally not true, and that the anticipation of the resetting event well before getting trapped is beneficial to navigate the network. In particular, for the cases studied, we observe that the cover time is minimized when we reset the memory when the walker is between the 8\% and the 50\% of its mean length.

\section{Conclusions}

All in all, we have shown that the presence of finite memory in the trajectories of self-avoiding random walkers yields an optimal value for the cover time, outperforming previously reported efficient strategies for network exploration, such as non-backtracking and pure self-avoiding walkers. We model this as a stochastic resetting problem, where a self-avoiding random walker repeatedly forgets its trajectory with a certain probability. We have analytically characterized the process via the reduction of the infinite memory of the pure self-avoiding random walk to a quasi-Markovian process of shrinking networks, where the main quantities at time $t$ depend only on those at time $t-1$ and $t-2$. Moreover, the method presented here is particularly convenient to give insights on large networks, since the analytical estimates based on degree distributions, which typically contain, at most, $k_{\text{max}} \sim \mathcal{O} (\sqrt N)$ values~\cite{boguna2004cut}, have a much lower algorithmic complexity than those of classical random walks, based on the graph's $N\times N$ adjacency matrix and its powers~\cite{masuda2017random}. By way of example, we can easily obtain analytical estimates for the position of the walker and for first-passage time and survival distributions in huge networks for which it would be unfeasible to use a matricial approach. Yet, the information provided by our framework comes aggregated by node degree, at odds with more traditional approaches, that comes at a single node level. Using one framework or another narrows down to balancing the granularity of the information we want to retrieve and the computational capabilities to calculate the quantities of interest.

By means of master and renewal equations, we have provided a characterization of the stochastic properties of our model at several scales. At the smallest scale, we have obtained the time-dependent probability to find the walker at a node of arbitrary degree. At the network level, we offer excellent estimates for the degree distribution of the non-visited network. These two metrics are the building blocks to address other quantities useful in the context of network search and navigability. Here, at the smallest scale we have the first passage to nodes of arbitrary degree, for which we observe a non-monotonic behavior with a global minimum, located in the area where the degree distribution attains its maximum, as one would expect. Remarkably, in the area of intermediate-to-large degrees, we report the emergence of a non-trivial local maximum, which our theory captures, consequence of the non-ergodic shrinking of the network during the navigation of the self-avoiding walker. At a larger scale, we discover an optimal value of the resetting parameter that minimizes the time to cover the network. In practical terms, we unravel the efficiency behind using finite memory in exploration scenarios. These results are expected to be robust, as far as the topological correlations in the explored network are not strong.

An important feature of our model is that it uses only local information. That is, the full topology of the network need not to be known, since the walker moves only to non-visited neighboring nodes. This locality, along with the addition of finite memory, offers some similarities and differences with well-known models in non-equilibrium statistical physics. Our model could be mapped to a version of run-and-tumble motion on networks, where the ``tumble'' events would correspond to memory erasures and the ``run'' events would correspond to the self-avoiding trajectories of the walker. Additionally, one could constrain the ``runs'' to moves that only step the walker further away from the departure node, yet this would significantly change the blocking conditions and would require an alternative mathematical treatment. Our model could be easily adapted to model some L{\'e}vy processes, that are known to be efficient exploration strategies under the right circumstances~\cite{kleinberg2000navigation, viswanathan1999optimizing}. In this context, when it comes to exploring networked systems, most works deal with the case of L{\'e}vy flights, i.e., walkers whose length jump is distributed according to a power law and whose transitions to any site occurs in single time step, no matter how far that site is~\cite{riascos2012long, weng2015levy, guo2016levy, estrada2018random, cipolla2021nonlocal}. These instantaneous propagation and discontinuity (i.e., nonlocality) in the trajectories might be unphysical in certain scenarios, specially when the walker is a physical entity or when the complete topology is not known. Our model better aligns with L{\'e}vy walks, where the walkers possess a finite velocity and long transitions take proportionally longer times; see, for instance,~\cite{dybiec2017levy} for a thorough discussion on the similarities and differences of L{\'e}vy flights and L{\'e}vy walks. Indeed, in our model the trajectories are always continuous (i.e., local) and the memory resetting trials at each time step with probability $r$ are equivalent to set a self-avoiding trajectory of length $\ell$, where $\ell$ is a random variable distributed according to a geometric distribution of parameter $r$, assuming the appropriate boundary conditions~\cite{dybiec2017levy}. Similarly, one could investigate trajectories whose length is distributed with an arbitrary probability density function, such as $p(\ell) \sim \ell^{-\gamma}$, with $\gamma \in (1, 3]$, typical of Levy processes. The renewal equations lend themselves to be generalized to arbitrary length distribution, that is, to arbitrary resetting statistics.

On these bases, we envision promising extensions of the mathematical framework presented here due to its flexibility, such as including more complex resetting statistics~\cite{evans2020stochastic}, modifying the dynamical rules underlying the walker moves, and considering more realistic network topologies, such as multilayer~\cite{artime2022multilayer} and higher-order networks~\cite{battiston2020networks}. We hope that our results become a stepping stone on the path toward improving network exploration, navigability and discovery, in particular, and toward disentangling the intricate interplay between memory and topology in complex systems, in general.

\section{Acknowledgments}
All authors thank Prof. V. Vinciotti for feedback at an intermediate stage of this work. G.C. acknowledges partial financial support from the International Office of the University of Trento. O.A. acknowledges financial support from the Spanish Ministry of Universities through the Recovery, Transformation and Resilience Plan funded by the European Union (Next Generation EU), and the University of the Balearic Islands.

\bibliography{biblio}

\pagebreak

\renewcommand{\thefigure}{SM\arabic{figure}}
\setcounter{figure}{0}
\renewcommand{\theequation}{SM\arabic{equation}}
\setcounter{equation}{0}
\widetext
\section*{Supplementary Materials for \\ \textit{Efficient network exploration by means of resetting self-avoiding random walkers}}

\begin{figure}[h!]
\includegraphics[width=.95\textwidth]{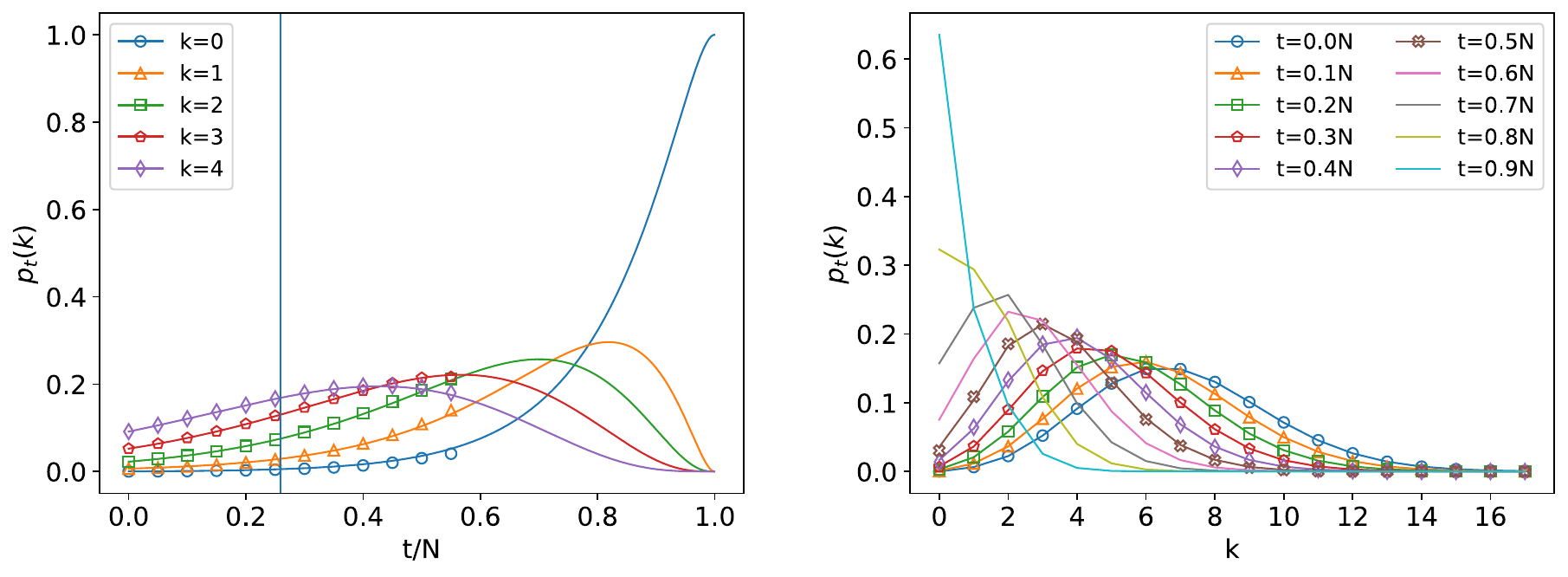}
\caption{\label{fig:deg_dist_er}SARW conditional temporal degree distribution for ER random graphs with $N=1000$ and $\langle k\rangle_0=7$. We compare our theoretical predictions (solid lines) with the simulation results, indicated by markers. (Left) The distribution $p_t(k)$ is represented as a function of $t/N$ for some fixed values of $k$. The vertical line represents the mean length of the SARW, that in this case is approximately equal to $170$. (Right) The distribution $p_t(k)$ is represented as a function of $k$ for some fixed values of $t$. The lack of markers for $t>0.5N$ in both panels is due to the absence of simulations surviving longer than $0.5N$. Simulations are run on a sample of $1000$ random networks from the ER model with fixed initial degree distribution, for 5 different initial conditions (randomly chosen initial nodes). Average values are shown.}
\end{figure}

\begin{figure}[h!]
\includegraphics[width=.95\textwidth]{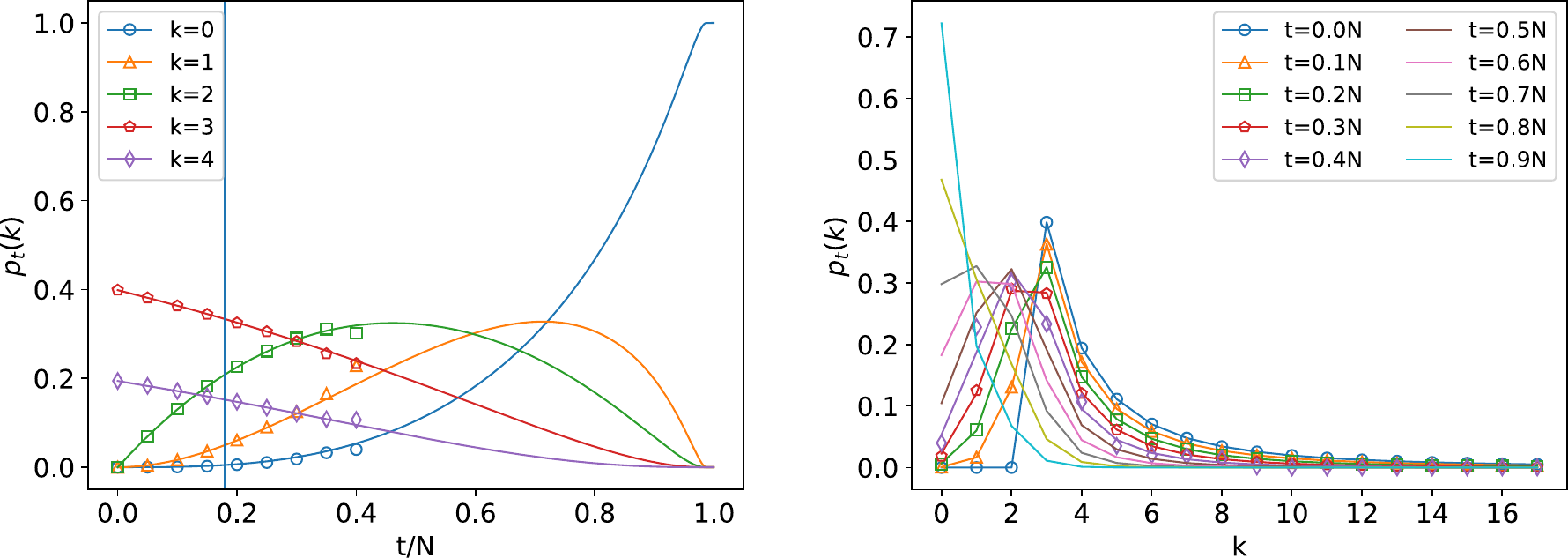}
\caption{\label{fig:deg_dist_sf}SARW conditional temporal degree distribution for SF random graph with $N=1000$ and $\alpha=2.5$. We compare our theoretical predictions (solid lines) with the simulation results, indicated by markers. (Left) The distribution $p_t(k)$ is represented as a function of $t/N$ for some fixed values of $k$. The vertical line represents the mean length of the SARW, that in this case is approximately equal to $120/N$. (Right) The distribution $p_t(k)$ is represented as a function of $k$ for some fixed values of $t$. The lack of markers for $t>0.4N$ in both panels is due to the absence of simulations surviving longer than $0.4N$. Simulations are run on a sample of $2000$ random networks from the SF model with fixed initial degree distribution. Since the degree distribution is more heterogeneous w.r.t. the ER model, here we chose 10 different initial conditions (randomly chosen initial nodes). Average values are shown.}
\end{figure}

\begin{figure}[h!]
\includegraphics[width=.95\textwidth]{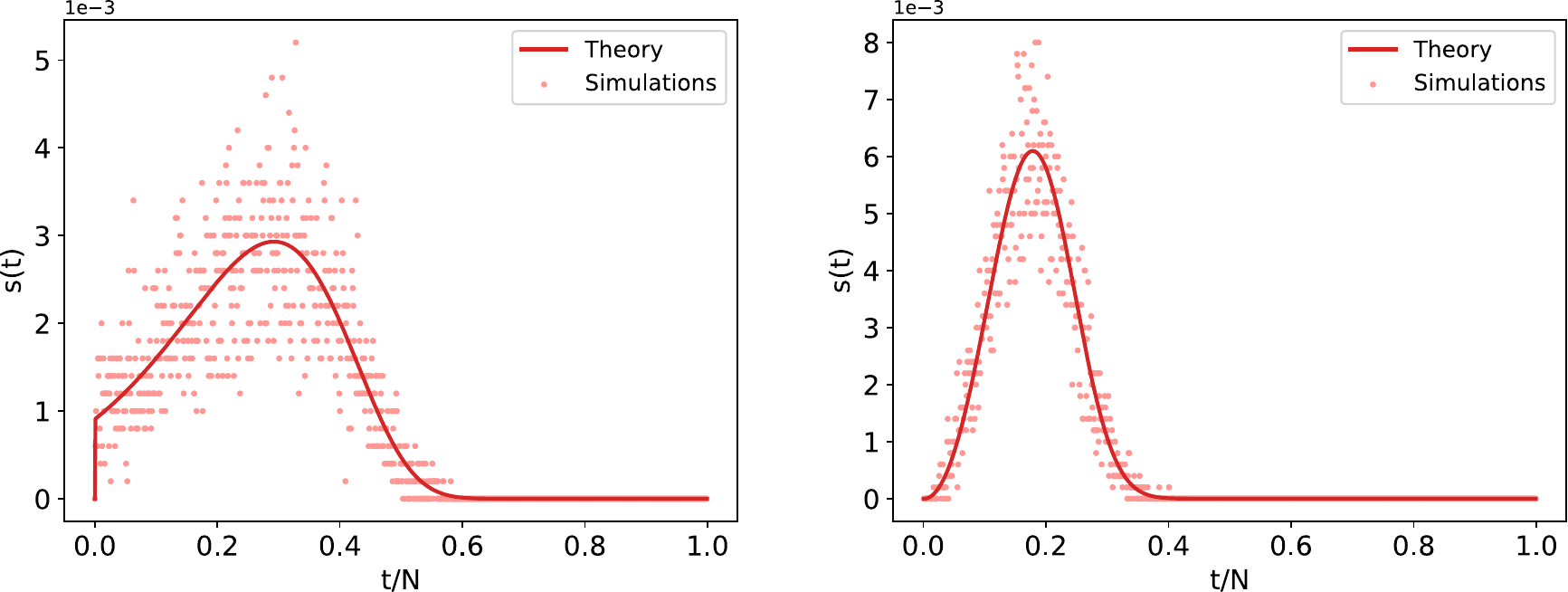}
\caption{\label{fig:stop_dist}SARW stopping distribution. We compare our theoretical predictions (solid lines) with the simulations results (dots). (Left) The distribution $s(t)$ is computed for ER random graph with $N=1000$ and $\langle k\rangle_0=7$. We simulated $5$ walks on $1000$ random networks with same initial degree distribution. (Right) The distribution $s(t)$ is computed for SF random graph with $N=1000$ and $\alpha=2.5$. Simulations are run on samples of $2000$ random networks from both topologies with fixed initial degree distribution, for $10$ different initial conditions (randomly chosen initial nodes). Average values are shown.}
\end{figure}

\begin{figure}[h!]
\includegraphics[width=.95\textwidth]{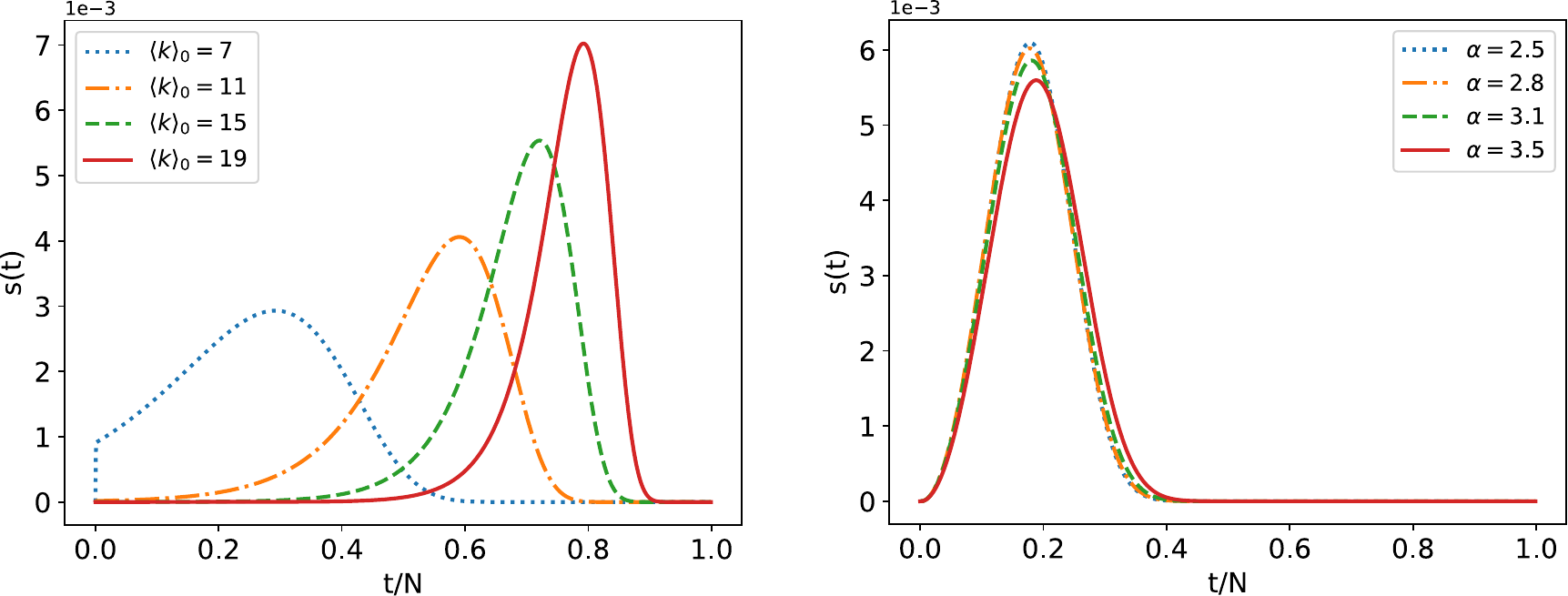}
\caption{\label{fig:stop_comp}SARW stopping distribution -- different topological parameter. We compare the theoretical predictions of the stopping distribution $s(t)$ for different topological parameters. (Left) $s(t)$ is computed for ER random networks with $N=1000$ and different initial mean degree $\langle k\rangle_0$, reported in the legend. The shape of $s(t)$ varies visibly for varying $\langle k\rangle_0$. This implies that also its moments, in particular the mean length of the SARW, change as functions of the initial average degree.(Right) $s(t)$ is computed for SF random networks with $N=1000$ and different exponent $\alpha$, reported in the legend. Here, varying the parameter $\alpha$ does not have an impact on the stopping distribution and, so, on the SARW's mean length. This dependency on the topological parameters agrees with what is reported in Fig.~\ref{fig:cov}.}
\end{figure}

\begin{figure}[h!]
\includegraphics[width=.95\textwidth]{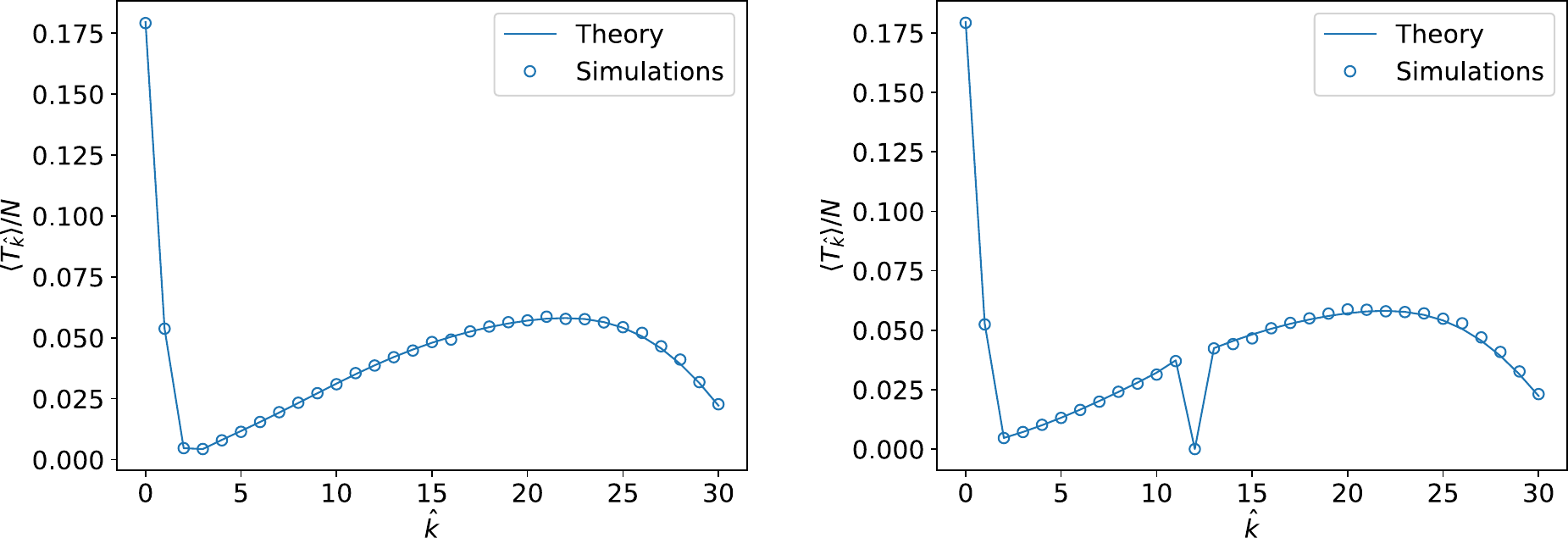}
\caption{\label{fig:SM_mfpt_sf}SARW on SF random graphs with $N=1000$ and $\alpha=2.5$ -- mean first passage time. We compare our theoretical predictions (solid lines) with the simulations results (markers). Simulations are run on a sample of $1000$ random networks from the SF model with fixed initial degree distribution, for 5 different initial conditions. (Left) The walks start on a node chosen uniformly at random. (Right) The walks start on a node chosen uniformly at random among those nodes of degree $12$.}
\end{figure}

\begin{figure}[h!]
\includegraphics[width=.95\textwidth]{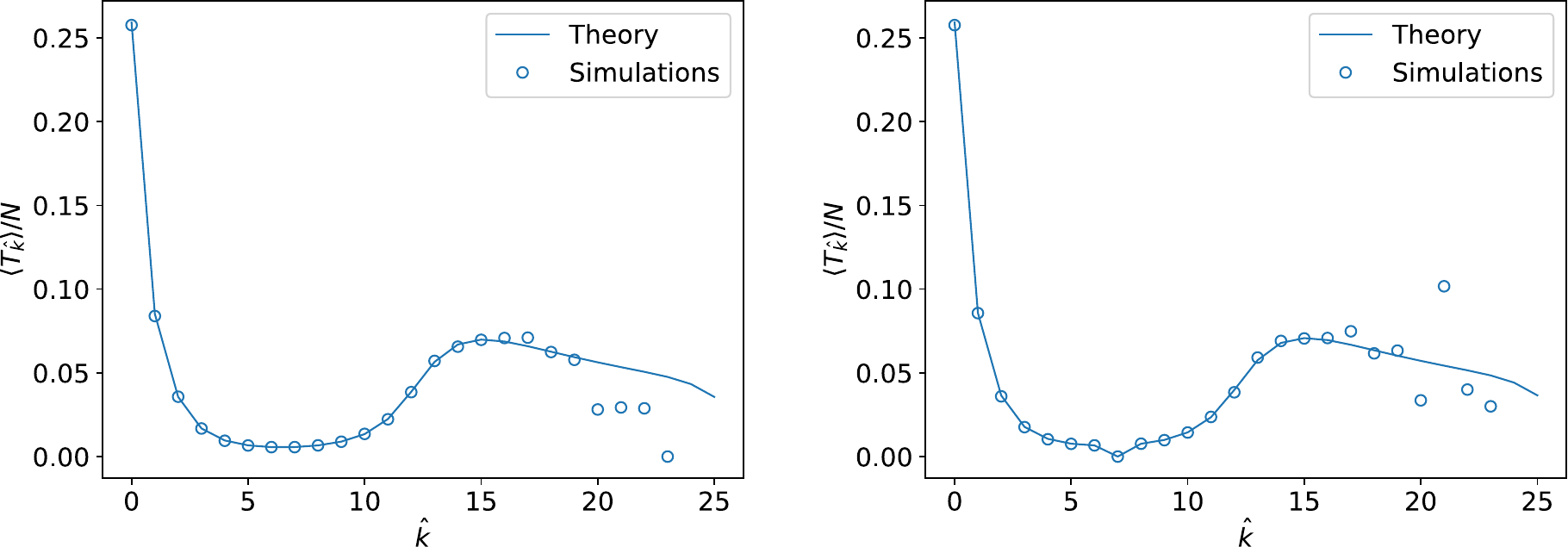}
\caption{\label{fig:SM_mfpt_er} SARW on ER random graphs with $N=1000$ and $\langle k\rangle_0=7$ -- mean first passage time. We compare our theoretical predictions (solid lines) with the simulations results (markers). Simulations are run on a sample of $1000$ random networks from the ER model with fixed initial degree distribution, for 5 different initial conditions. (Left) The walks start on a node chosen uniformly at random. (Right) The walks start on a node chosen uniformly at random among those nodes having degree $7$. The fluctuations for the high values of $\hat{k}$ are due to the fact that only for few realizations of the process, the walker reaches those degrees before reaching a node of degree $0$. Recall that the degree distribution of ER random graphs is approximately Poisson, so that $\langle k\rangle_0=7$ represents both the mean and variance of the initial degree distribution. At $\hat{k} = 20$, we are about five standard deviations away from the average degree.}
\end{figure}

\begin{figure}[h!]
\includegraphics[width=.95\textwidth]{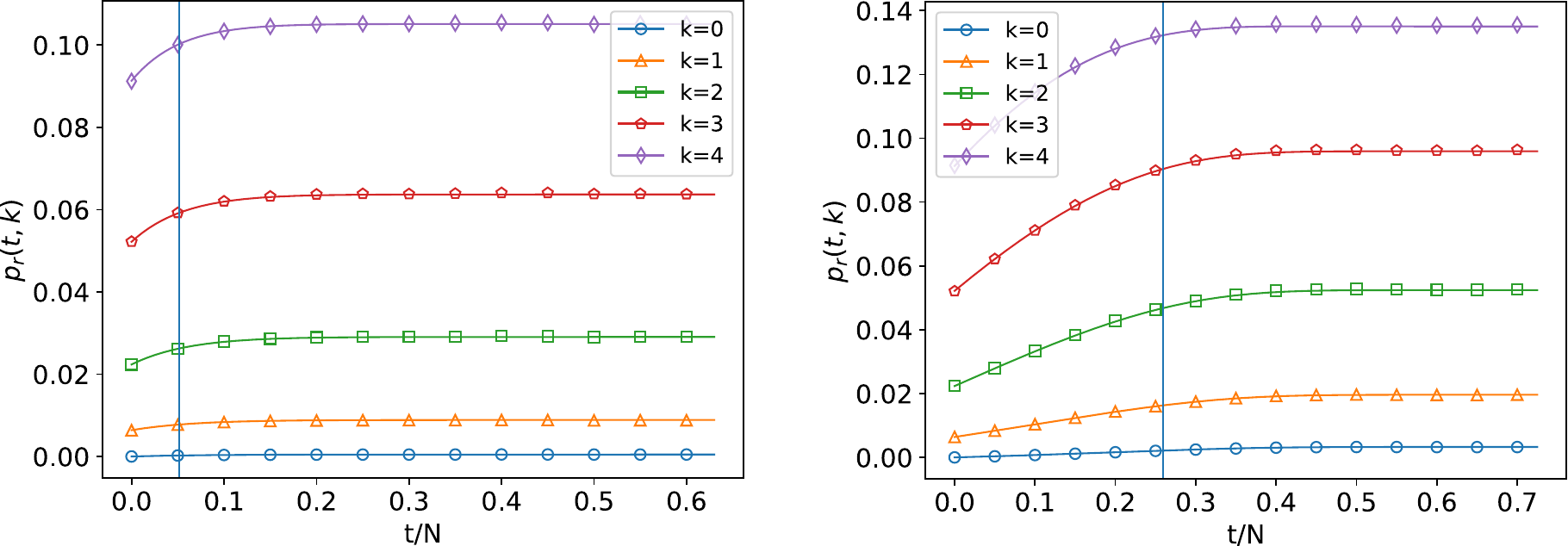}
\caption{\label{fig:pr(tk)_ER}SARWR temporal degree distribution for ER random graphs with $N=1000$, $\langle k\rangle_0=7$ and different resetting parameter $r$. We compare our theoretical predictions (solid lines) with the simulation results, indicated by markers. The distribution $p_r(t,k)$ is represented as a function of time for some fixed values of $k$. Simulations are run on a sample of $1000$ random networks from the ER model with fixed initial degree distribution, for 5 different initial conditions (randomly chosen initial nodes). Average values are shown. (Left) The resetting parameter $r$ is approximately equal to $1/51$. (Right) The resetting parameter $r$ is approximately equal to $1/259$. The vertical lines represent the $\frac{1}{r}$ value.}
\end{figure}

\begin{figure}[h!]
\includegraphics[width=.95\textwidth]{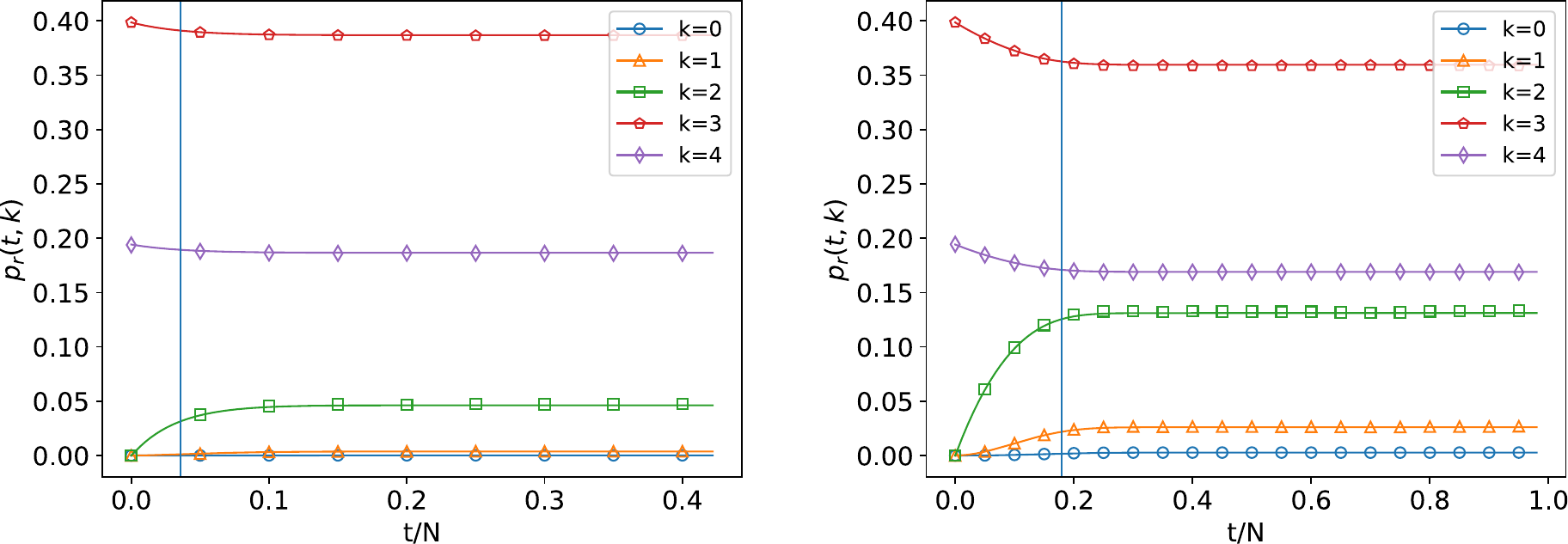}
\caption{\label{fig:pr(tk)_SF}SARWR temporal degree distribution for SF random graphs with $N=1000$, $\alpha=2.5$ and different resetting parameter $r$. We compare our theoretical predictions (solid lines) with the simulation results, indicated by markers. The distribution $p_r(t,k)$ is represented as a function of time for some fixed values of $k$. Simulations are run on a sample of $1000$ random networks from the SF model with fixed initial degree distribution, for 5 different initial conditions (randomly chosen initial nodes). Average values are shown. (Left) The resetting parameter $r$ is approximately equal to $1/35$. (Right) The resetting parameter $r$ is approximately equal to $1/179$. The vertical lines represent the mean resetting time $\frac{1}{r}$.}
\end{figure}

\begin{figure}[h!]
\includegraphics[width=.95\textwidth]{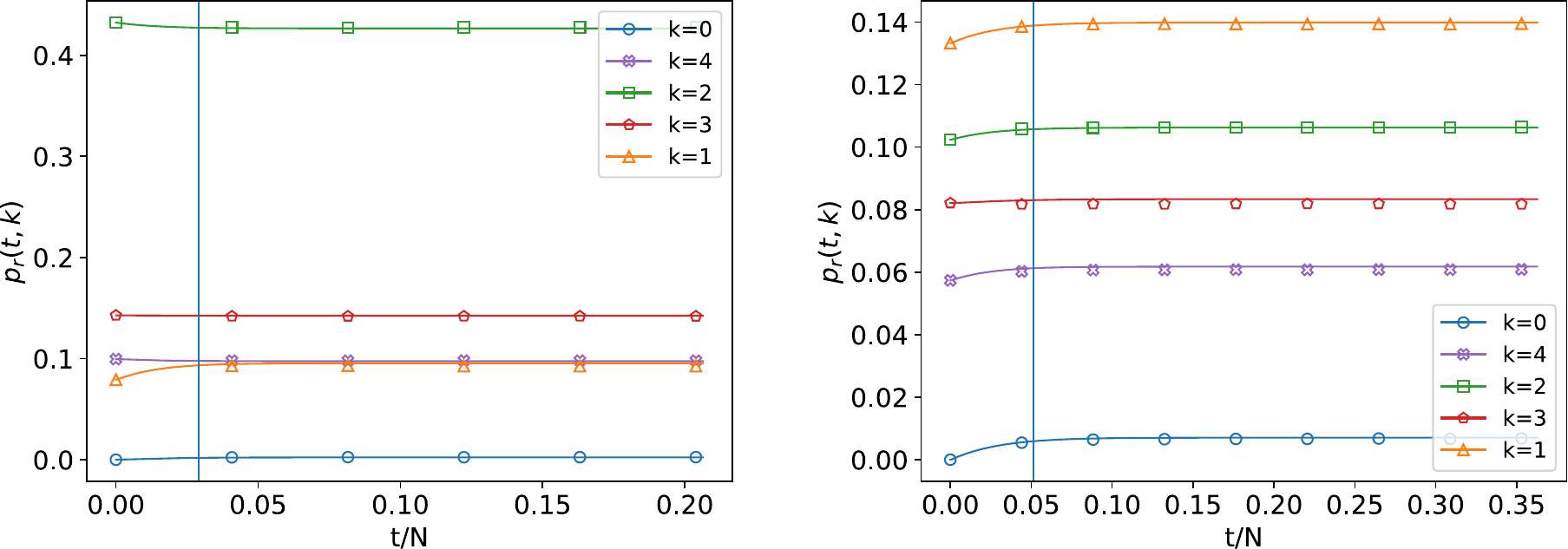}
\caption{\label{fig:pr(tk)_realnets}SARWR temporal degree distribution for the two real networks. We compare our theoretical predictions (solid lines) with the simulation results, indicated by markers. The distribution $p_r(t,k)$ is represented as a function of time for some fixed values of $k$. Simulations are run for $5000$ different initial conditions (randomly chosen initial nodes). Average values are shown. (Left) $p_r(t,k)$ is computed for the FAA network, and the resetting parameter $r$ is approximately equal to $1/35$. (Right) $p_r(t,k)$ is computed for the email's network, and the resetting parameter $r$ is approximately equal to $1/58$. The vertical lines represent the mean resetting time value, $\frac{1}{r}$, that in these plots corresponds to the mean average length of the SARW on each network.}
\end{figure}

\begin{figure}[h!]
\includegraphics[width=.95\textwidth]{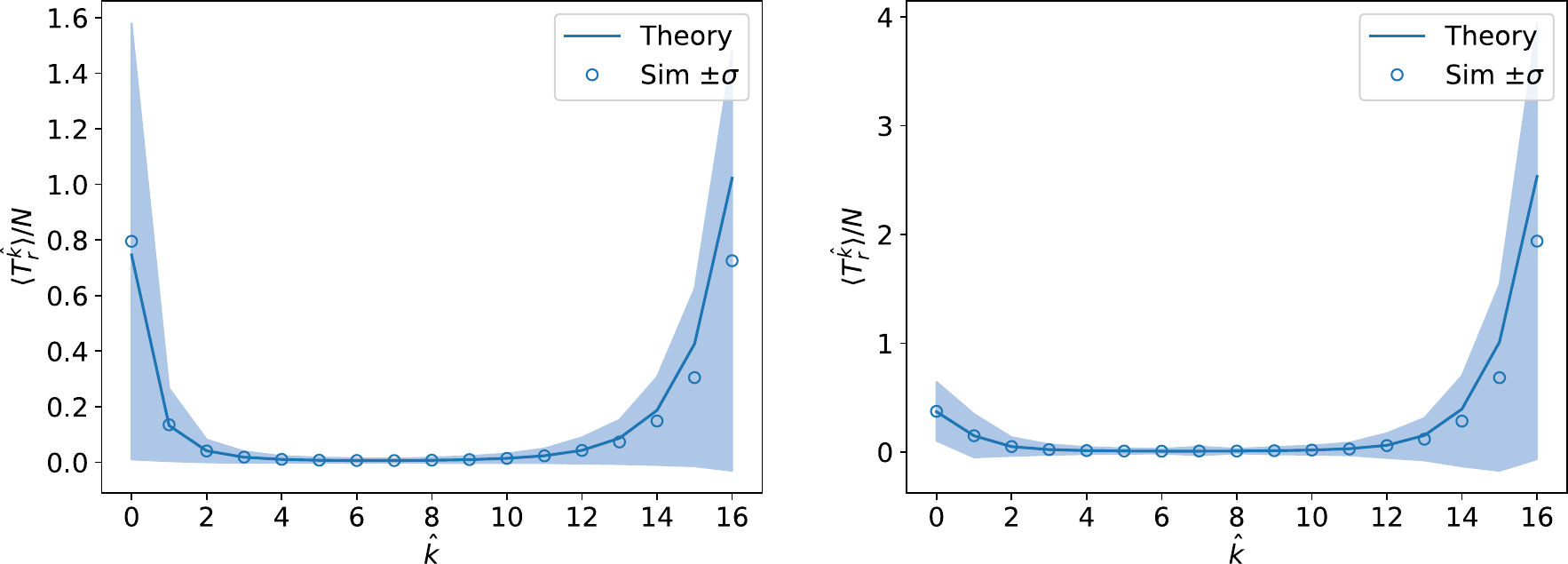}
\caption{\label{fig:mfpt_res_er}SARWR on ER random graphs with $N=1000$ and $\langle k\rangle_0=7$ -- mean first passage time. We compare our theoretical predictions (solid lines) with the simulations results (markers), for two values of the resetting parameter: $r\approx1/51$ (left)and $r \approx 1/259$ (right). The light-blue color illustrates the standard deviation. Simulations are run on a sample of $5000$ random networks from the ER model with fixed initial degree distribution. }
\end{figure}

\begin{figure}[h!]
\includegraphics[width=.95\textwidth]{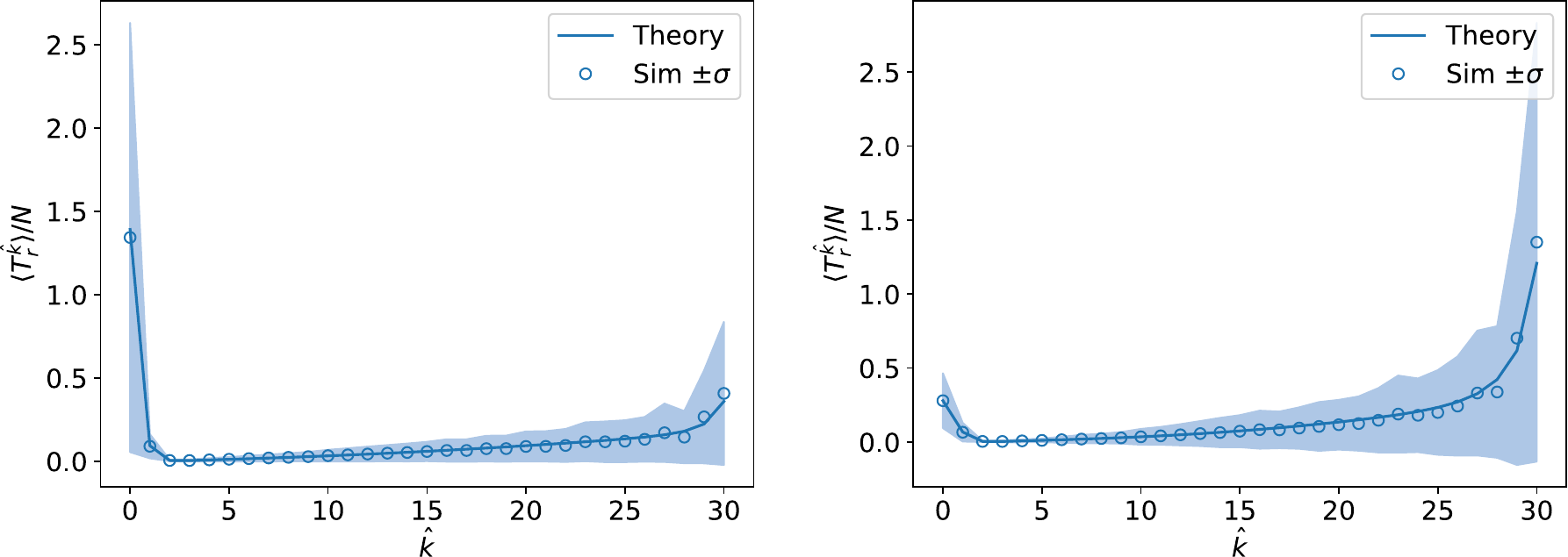}
\caption{\label{fig:mfpt_res_sf}SARWR on SF random graphs with $N=1000$ and $\alpha=2.5$ -- mean first passage time. We compare our theoretical predictions (solid lines) with the simulations results (markers), for two values of the resetting parameter: $r\approx1/35$ (left) and $r \approx 1/179$ (right). The light-blue color illustrates the standard deviation. Simulations are run on a sample of $5000$ random networks from the SF model with fixed initial degree distribution.}
\end{figure}

In this supplemental material file we provide the following content, that supports the results discussed in the main article:
\begin{itemize}
    \item details on the derivation of the recursive equation for the time-dependent degree distribution $p_t(k)$ of the network of non-visited nodes by a self-avoiding random walker;
    \item details on the synthetic and empirical networks that we employ for our analyses;
    \item additional figures for the analysis of the self-avoiding random walk, namely, the degree distributions $p_t(k)$ (Figs.~\ref{fig:deg_dist_er} and \ref{fig:deg_dist_sf}) and $p(t,k)$ (Fig.~\ref{fig:p(t,k)}), the survival probability $s(t)$ (Figs.~\ref{fig:stop_dist} and \ref{fig:stop_comp}) and the mean first-passage time $\langle T_{\hat{k}} \rangle$ to a target $\hat{k}$ (Figs.~\ref{fig:SM_mfpt_sf} and \ref{fig:SM_mfpt_er});
    \item additional figures for the analysis of the self-avoiding random walk with memory, namely, the degree distribution $p_r(t,k)$ (Figs.~\ref{fig:pr(tk)_ER}, \ref{fig:pr(tk)_SF}, \ref{fig:pr(tk)_realnets}), the mean first-passage time $\langle T_r^{\hat{k}} \rangle$ to a target $\hat{k}$ (Figs.~\ref{fig:mfpt_res_er} and \ref{fig:mfpt_res_sf}), the time-dependent position of the walker $d_r(t,k)$ (Figs.~\ref{fig:dr(t,k)_er}, \ref{fig:dr(t,k)_sf} and \ref{fig:dr(t,k)_realnets});
\end{itemize}

\subsection*{Self-avoiding random walk (SARW) -- Time-dependent degree distribution}

Let us begin with the time-dependent degree distribution $p_t(k)$. Firstly, to obtain the formula that captures the evolution of the difference of nodes of degree $k$ in a single time step $D_t(k|0^c)$, see Eq.~\eqref{eq:Dt}, we have to consider the three cases, similar to what is done in Ref.~\cite{valente2022non}:
\begin{itemize}
    \item [(a)] The node deleted at time $t-1$ could be of degree $k$, and this happens with probability $q_{t-2}(k)$. Thus, on average, $q_{t-2}(k)$ degree-$k$ nodes are deleted.
    \item [(b)] The degree-$(k + 1)$ neighbors of the node visited at time $t-1$ lose an edge and become of degree $k$. Let us denote by $N_{t-1}(k+1|s)$ the number of nodes of degree $k+1$, which are neighbors of a degree-$s$ node. Then, summing over all the possibilities, we obtain the number of degree-$(k+1)$ neighbors of the deleted node
    \begin{equation}
        \sum_{s=0}^{N_t}q_{t-2}(s)N_{t-1}(k+1|s)\approx\sum_{s=0}^{N_t}q_{t-2}(s)sq_{t-1}(k)=\langle r\rangle_{t-2}q_{t-1}(k);
    \end{equation}
    \item[(c)] the degree-$k$ neighbors of the node visited at time $t-1$ lose an edge and become of degree $k-1$. Let's denote with $N_{t-1}(k|s)$ the number of nodes of degree $k+1$ neighbors of a degree-$s$ node. Then, summing over all the possibilities, we obtain the number of degree-$k$ neighbors of the deleted node
    \begin{equation}
        \sum_{s=0}^{N_t}q_{t-2}(s)N_{t-1}(k|s)\approx\sum_{s=0}^{N_t}q_{t-2}(s)sq_{t-1}(k-1)=\langle r\rangle_{t-2}q_{t-1}(k-1).
    \end{equation}
\end{itemize}
Both for cases (b) and (c) the approximation is justified by the assumption that the nodes' degrees  in the network are uncorrelated, so that
$N_{t-1}(k+1|s)$ is approximately equal to $s$ times the probability that a neighbor of a randomly selected node has degree $k+1$ (or equivalently, excess degree $k$) that is $q_{t-1}(k)$. 
Putting the three terms together, we obtain Eq.~\eqref{eq:Dt}
\begin{equation}\tag{\ref{eq:Dt}}
    D_t(k|0^c)\approx\langle r\rangle_{t-2}[q_{t-1}(k)-q_{t-1}(k-1)]-q_{t-2}(k).
\end{equation}
Let us, then, focus on the time-dependent degree distribution, given by Eq.~\eqref{eq:ptk-node}. 
We have to consider two events: the walker stops at time $t-1$ and it happens with probability $q_{t-2}(0)$, or it does not, which happens with probability $1-q_{t-2}(0)$. 
Then, by the law of total probability, we obtain 
\begin{align}
    p_t(k)&=\dfrac{N_t(k)}{N_t}=\dfrac{N_{t-1}(k)+D_t(k)}{N_t} = q_{t-2}(0)\left[\dfrac{N_{t-1}(k)+D_t(k|0)}{N_{t-1}}\right]+(1-q_{t-2}(0))\left[\dfrac{N_{t-1}(k)+D_t(k|0^c)}{N_{t}}\right]\notag\\
    &= q_{t-2}(0)p_{t-1}(k)+(1-q_{t-2}(0))\left[\dfrac{N_{t-1}(k)+D_t(k|0^c)}{N_{t}}\right]\tag{\ref{eq:ptk-node}}
\end{align}
and, finally, using Eq.~\eqref{eq:ptk-node} we can evaluate an approximated $p_t(k)$.
Given this approximation, $p_t$ may not be a proper distribution.
Let us suppose $p_{t-1}$, $q_{t-1}$ and $q_{t-2}$ are normalized and let us compute the normalization constant for $p_t$
\begin{align*}
    C_t&=\sum_{k=0}^{N-1}p_t(k)=q_{t-2}(0)+(1-q_{t-2}(0))\left[\dfrac{N_{t-1}+\sum_k D_t(k|0^c)}{N_t}\right]\\
    &=q_{t-2}(0)+\dfrac{1-q_{t-2}(0)}{N_t}\left[N_{t-1}+\langle r\rangle_{t-2}q_{t-1}(0)+\sum_{k=1} D_t(k|0^c)\right]\\
    &=q_{t-2}(0)+\dfrac{1-q_{t-2}(0)}{N_t}\\&\left[N_{t-1}+\langle r\rangle_{t-2}q_{t-1}(0)-(1-q_{t-2}(0))+\langle r \rangle_{t-2}[(1-q_{t-1}(0))-(1-q_{t-2}(N-1))]\right]\\
    &=q_{t-2}(0)+\dfrac{1-q_{t-2}(0)}{N_t}\left[N_{t-1}+\langle r\rangle_{t-2}-(1-q_{t-2}(0))+\langle r \rangle_{t-2}[-1+q_{t-2}(N-1)]\right]\\
    &=q_{t-2}(0)+\dfrac{1-q_{t-2}(0)}{N_t}\left[N-t+1-1+q_{t-2}(0)+\langle r \rangle_{t-2}q_{t-2}(N-1)\right]\\
    &=q_{t-2}(0)+(1-q_{t-2}(0))\left[1+\dfrac{q_{t-2}(0)+\langle r \rangle_{t-2}q_{t-2}(N-1)}{N_t}\right]\\
    &=1+\dfrac{(1-q_{t-2}(0))(q_{t-2}(0)+\langle r \rangle_{t-2}q_{t-2}(N-1))}{N_t}=1+\dfrac{(1-q_{t-2}(0))q_{t-2}(0)}{N_t}.
\end{align*}
Observe that the support of $p_t$ depends on $t$ since, as time passes the probability of some degrees becomes null. Nevertheless, we can always add zeros to our summation and simply write $\sum_{k = 0}^{N-1} \cdot$.
The normalization constant differs from $1$ by a quantity that depends on $N_t$ and $q_{t-2}(0)$. For small $t$ (or, equivalently, large $N_t$) we have
\begin{equation*}
    \dfrac{(1-q_{t-2}(0))q_{t-2}(0)}{N_t}\le\dfrac{1}{N_t}\approx 0.
\end{equation*}
Instead, for large $t$, the probability to reach a degree-$0$ node goes to $1$, and the last term is still very close to zero, because the numerator goes to zero.
In conclusion, we take Eq.~\eqref{eq:ptk-node} as a good approximation of a distribution.\\

\subsection*{Network details}

All the theoretically derived equations presented in the main article have been extensively tested via numerical simulation. 
We considered Erd{\H{o}}s-R{\'{e}}nyi (ER) random graphs with initial degree distribution
\begin{gather*}
    p_0(k)\propto\dfrac{\langle k\rangle_0^k}{k!}e^{-\langle k\rangle_0}\qquad 0<k\le k_{\text{max}},\\
    \text{with } k_{\text{max}} \text{ such that } \sum_{k=k_{\text{max}}}^{+\infty}\dfrac{\langle k\rangle^k}{k!}e^{-\langle k\rangle}<\dfrac{1}{N},
\end{gather*}
where $\langle k\rangle_0=7,11,15,19$ and $N=1000,2200,4600$ and $10000$.
For the same values of $N$, we considered also scale-free (SF) random network with initial degree distribution \begin{equation*}
    p_0(k)\propto k^{-\alpha}\qquad 3\le k\le\lfloor \sqrt{N}\rfloor.
\end{equation*}
where $\alpha=2.5,2.8,3.1$ and $3.5$. Here, we present the results only for ER with $N=1000$ and $\langle k\rangle_0=7$, and SF with $N=1000$ and $\alpha=2.5$.

Self-avoiding random walks with resetting are studied in the synthetic networks mentioned above, as well as in two empirical networks.
The first one represents the exchange of emails among members of the Rovira i Virgili University in Catalonia, in 2003 \cite{guimera2003self}. 
Nodes represent email addresses, and two nodes are linked if they exchanged at least an email, with reply (unidirectional links were disregarded).
Emails sent to more than $50$ recipients were not considered.  
We consider only the largest connected component, which consists of $1133$ nodes.
The degree distribution is fitted well by an exponential $p_0(k)\propto\exp(-k/k^*)$, for $k\ge2$ and $k^*=9.2$.
The empirical degrees range from 1 to 34, with an average of, approximately, 9.6.
The second real network has been constructed by the USA's FAA (Federal Aviation Administration) \cite{kunegis2013konect,Maayan2006} and is henceforth referred to as the FAA network.
It has $1226$ nodes, representing the airports or service centers. Links are created from strings of preferred routes recommended by the NFDC (National Flight Data Center). 
A power-law $p_0(k)\propto x^{-\alpha}$, with $\alpha=1.9$ fits well its degree distribution.
Here, the (empirical) average degree is 3.9, while the maximum is 71. Since 95.5\% of the nodes have degree at most 30, we restrict our computations, e.g., for the mean first-passage time, to $k=30$ for both real networks. 

\subsection*{Self-avoiding random walk (SARW) -- Additional figures}

\begin{figure}[h!]
\includegraphics[width=.95\textwidth]{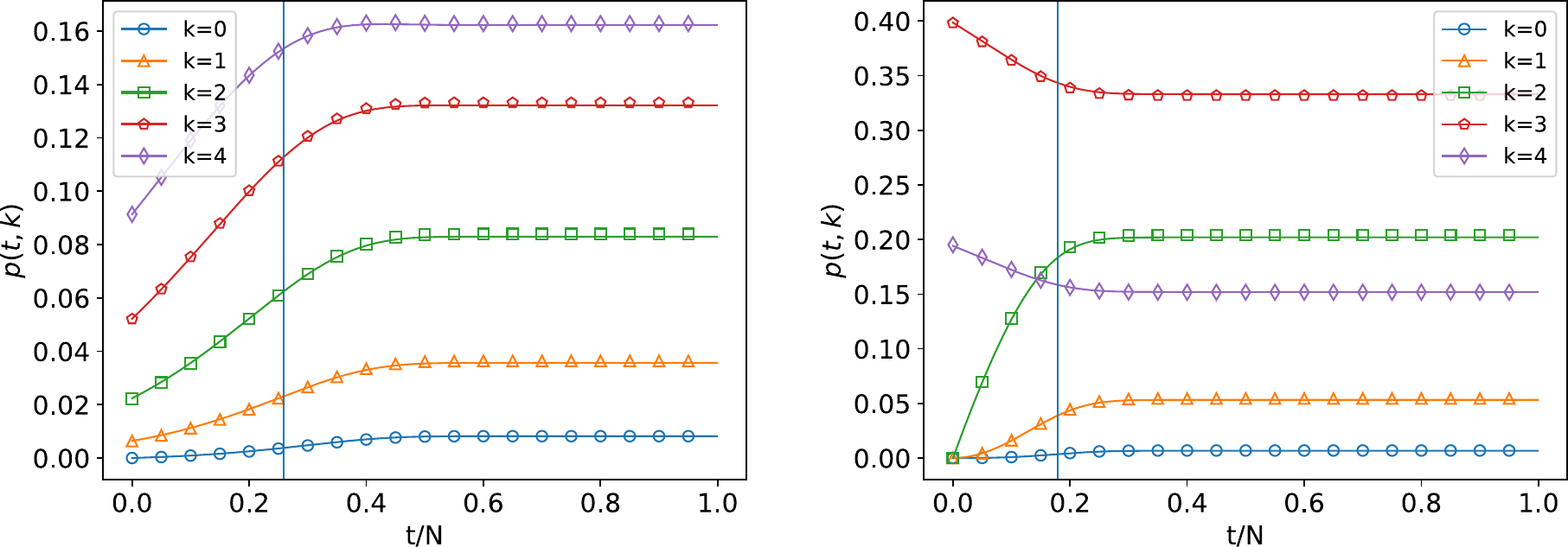}
\caption{\label{fig:p(t,k)}SARW temporal degree distribution. We compare our theoretical predictions (solid lines) with the simulation results, markers. The distribution $p(t,k)$ is represented as a function of time for some fixed values of $k$. (Left) Distribution for ER random graphs with $N=1000$ and $\langle k\rangle_0=7$; simulations are run on a sample of 1000 random networks with fixed initial degree distribution, for 5 different initial conditions (randomly chosen initial nodes). (Right) Distribution for SF random graphs with $N=1000$ and $\alpha=2.5$; simulations are run on a sample of 2000 random networks with fixed initial degree distribution, for 10 different initial conditions (randomly chosen initial nodes). Average values are shown. The vertical lines represent the mean length of the SARWs on both random topologies, which are approximately $170/N$ and $120/N$, respectively.}
\end{figure}

\clearpage

\subsection*{Self-avoiding random walk with resetting (SARWR) -- Additional figures}

Let us first provide more details on the resetting parameter $r$. Since it is reasonable to assume that this parameter is intrinsic to the walking process, we chose to keep $r$ constant across the walk, i.e., $r (t) \equiv r$.
The probability that the first resetting takes place at time $t$ is described by a geometric random variable $R_1$ with parameter $r$, with distribution
\begin{align*}
    \mathbb{P}(R_1=t) & =(1-r)^{1-t}r,\\
    \mathbb{E}[R_1] & =\dfrac{1}{r}.
\end{align*}
Considering identically distributed and independent consecutive resettings $R_i$, $\mathbb E[ R_i]$ also represents the mean waiting time ($\tau$) between two consecutive resetting events. 
Values of $r$ close to 1 lead to the classical random walk, while for $r \to 0$ we are, again, in the SARW scenario. 
We observed that an interesting range for the resetting parameter is $[\langle L \rangle ^ {-1}, 1]$, since $r^{\text{opt}}$ is always contained in it. This aspect can be easily seen in Fig. \ref{fig:cov}, noticing that on the $x$-axis we represent the values $\tau/\langle L\rangle$ that correspond to $[r\langle L\rangle]^{-1}$.

\begin{figure}[h!]
\includegraphics[width=.95\textwidth]{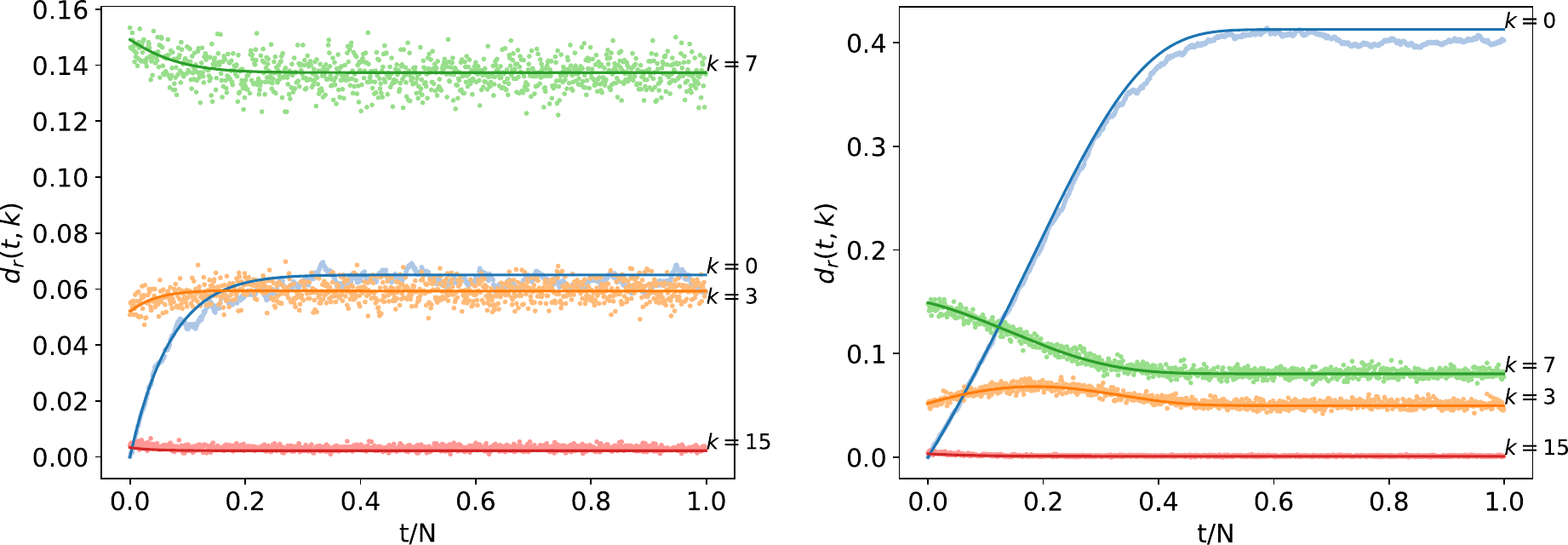}
\caption{\label{fig:dr(t,k)_er} SARWR on ER random graphs with $N=1000$ and $\langle k\rangle_0=7$ -- probability to find the walker on a degree-$k$ node. We compare our theoretical predictions (solid lines) with the simulations results (markers). The panels show the probability $d_r(t,k)$ as a function of time for some fixed value of $k$ and different values of the resetting parameter $r$. The values on the left panel have been computed with $r\approx1/51$. On the right panel, $r$ was fixed approximately equal to $1/259$. Simulations are run on a sample of $5000$ random networks from the ER model with fixed initial degree distribution. Average values are shown.}
\end{figure}

\begin{figure}[h!]
\includegraphics[width=.95\textwidth]{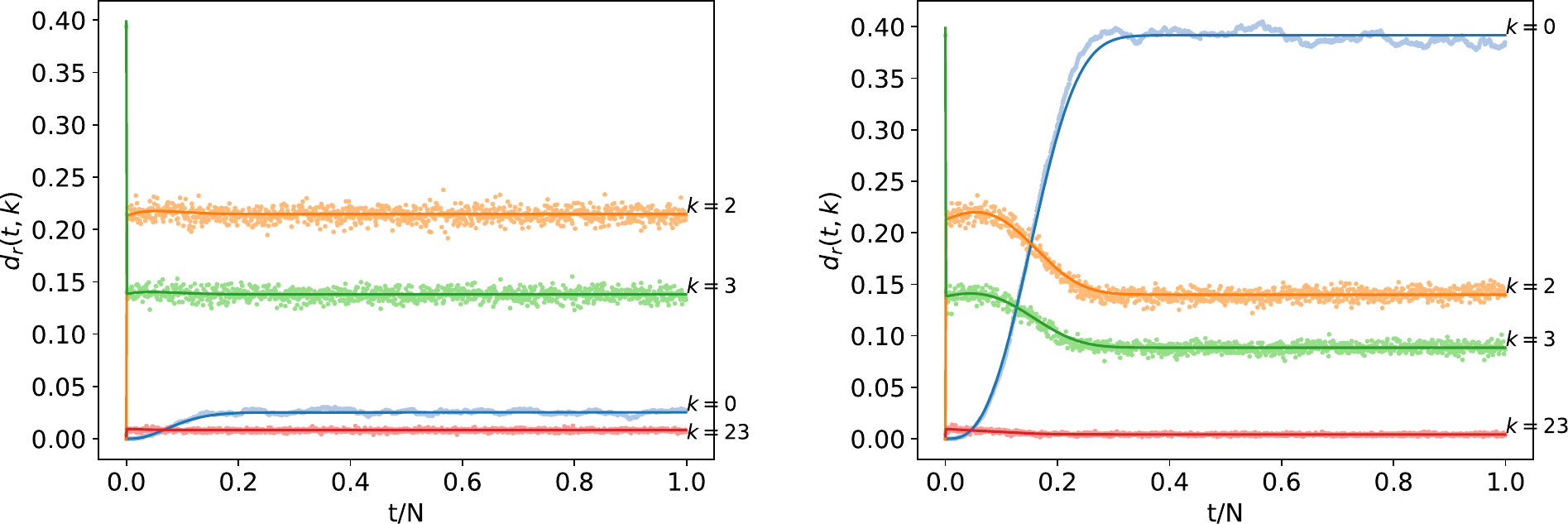}
\caption{\label{fig:dr(t,k)_sf}SARWR on SF random graphs with $N=1000$ and $\alpha=2.5$ -- probability to find the walker on a degree-$k$ node. We compare our theoretical predictions (solid lines) with the simulations results (markers). The panels show the probability $d_r(t,k)$ as a function of time for some fixed value of $k$ and different values of the resetting parameter $r$. The values on the left panel have been computed with $r\approx1/35$. On the right panel, $r$ was fixed approximately equal to $1/179$. Simulations are run on a sample of $5000$ random networks from the SF model with fixed initial degree distribution. Average values are shown.}
\end{figure}

\begin{figure}[h!]
\includegraphics[width=.95\textwidth]{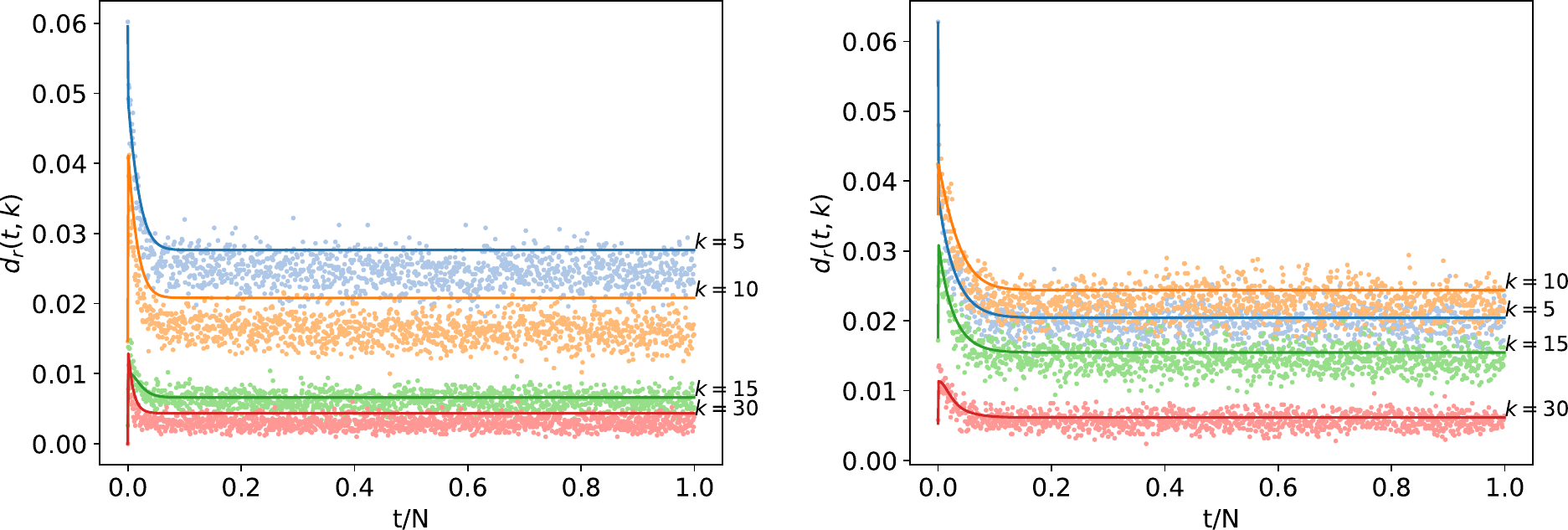}
\caption{\label{fig:dr(t,k)_realnets} SARWR on the two real networks -- probability to find the walker on a degree-$k$ node. We compare our theoretical predictions (solid lines) with the simulations results (markers). The panels show the probability $d_r(t,k)$ as a function of time for some fixed value of $k$. (Left) $d_r(t,k)$ is computed for the FAA network and the resetting parameter $r$ is approximately equal to $1/35$. (Right) $d_r(t,k)$ is computed for the email's network and the resetting parameter $r$ is approximately equal to $1/58$. Simulations are run for $5000$ different initial conditions (randomly chosen initial nodes). Average values are shown.}
\end{figure}

\end{document}